\documentclass[10pt,showpacs,nofootinbib]{revtex4}
\usepackage{amsmath,amssymb,amsfonts,graphicx}
\usepackage{tikz}  

\newcommand{\g}{{\mathfrak{g}}}
\def\half{\frac{1}{2}} 

\def\ben{\begin{equation}}
\def\een{\end{equation}}
\def\bea{\begin{eqnarray}} 
\def\eea{\end{eqnarray}}
\newcommand{\sect}[1]{\setcounter{equation}{0}\section{#1}}

\def\M{\mathcal{M}}

\def\nn{\nonumber}
\def\bR{\mathbb{R}}
\def\Tr{{\rm Tr }}
\def\ie{{\it i.e.\ }}
\def\eg{{\it e.g.\ }}

\input amssym.def
\input amssym.tex

\def\extd{\mathrm {d}}
\newcommand{\su}{\mathfrak{su}}

\newcommand{\SU}{\mathrm{SU}}

\newcommand{\R}{\mathbb{R}}

\newcommand{\be}{\begin{equation}}
\newcommand{\ee}{\end{equation}}
\newcommand{\bes}{\begin{eqnarray}}
\newcommand{\ees}{\end{eqnarray}}

\def\la{\langle}
\def\ra{\rangle}
\def\vphi{\varphi}

\begin{document}
\begin{flushright}AEI-2011-84\end{flushright}
\title{Discrete and continuum third quantization of Gravity}

\author{Steffen Gielen\footnote{\tt gielen@aei.mpg.de}, Daniele Oriti\footnote{\tt doriti@aei.mpg.de}}
\affiliation{Max Planck Institute for Gravitational Physics (Albert Einstein Institute), Am M\"uhlenberg 1, D-14476 Golm, Germany, EU} 

\begin{abstract}
We give a brief introduction to matrix models and the group field theory (GFT) formalism as realizations of the idea of a third quantization of gravity, and present in some more detail the idea and basic features of a continuum third quantization formalism in terms of a field theory on the space of connections, building up on the results of loop quantum gravity that allow to make the idea slightly more concrete. We explore to what extent one can rigorously define such a field theory. Concrete examples are given for the simple case of Riemannian GR in 3 spacetime dimensions. We discuss the relation between GFT and this formal continuum third quantized gravity, and what it can teach us about the continuum limit of GFTs. 
\end{abstract}

\pacs{04.60.Ds, 04.60.Nc, 04.60.Pp}

\maketitle

\sect{Introduction}
Several competing approaches to quantum gravity have been developed over the last years \cite{libro}, with a remarkable convergence, in terms of mathematical structures used and basic ideas shared. The group field theory (GFT) approach \cite{iogft} seems to lie at the crossroads of the canonical loop quantum gravity approach, its covariant spin foam counterpart, simplicial quantum gravity, matrix models and non-commutative geometry.  Like most of the other approaches, as a candidate description of the microscopic structure of quantum space in terms of discrete building blocks, the most important open issue that group field theory still faces is that of the ``continuum,'' \ie to show that a continuum spacetime emerges in a suitable approximation of the quantum discrete structures it is based on, that geometric fields emerge as the appropriate continuum variables in this regime and that their dynamics is well approximated then by classical General Relativity. Of course, this problem takes different forms, and implies different strategies for its resolution, depending on the mathematical and conceptual framework one uses. The one that is appropriate for group field theory seems to be that of ``third quantized gravity,'' in the sense that the naive analogue of group field theory in the context of continuum spacetime is a third quantization formalism in connection variables, in the same sense in which loop quantum gravity is the connection formalism that replaces continuum quantum geometrodynamics in ADM variables. Let us clarify a bit further.

\

Group field theories have been proposed \cite{laurentgft,iogft,iogft2,ProcCapeTown} as a kind of second quantization of canonical loop quantum gravity, in a discrete context, in the sense that one turns into a dynamical (quantized) field the canonical wave function of loop quantum gravity, associated to graphs embedded in space. From a related perspective, they can  be also understood as second quantizations, \ie field theories, of simplicial geometry \cite{iogft2}, or of the wave function of any canonical formulation of simplicial quantum gravity. Because both canonical loop quantum gravity and canonical simplicial gravity are supposed to represent the quantization of a (classical) field theory, General Relativity, this further quantization step brings GFT into the conceptual framework of ``third quantization'', a rather inappropriate label for a rather appealing idea: a field theory on the space of geometries, rather than spacetime, which also allows for a dynamical description of topology change. This idea has been brought forward more than 20 years ago in the context of canonical geometrodynamics, but has never been seriously developed due to huge mathematical difficulties, nor recast even formally in the language of connection dynamics, which allowed lately so much progress in the loop quantum gravity approach. We do so to some extent in this paper.

\

In this contribution, we give a brief introduction to the group field theory formalism, and we present in some more detail the idea and basic features of a continuum third quantization formalism in the space of connections, building up on the results of loop quantum gravity that allow to make the original idea slightly more concrete. We explore to what extent one can rigorously define such a field theory. Where concrete examples are given, we shall restrict ourselves to the simple case of Riemannian GR in 3 spacetime dimensions. Finally, we discuss the relation between GFT and this formal continuum third quantized gravity, and what it can teach us about the continuum limit of GFTs.

\ 

Our purpose (and, hopefully, the value of this contribution) is threefold: partly pedagogical, partly technical, partly motivational. 

The third quantized framework in the continuum is not a well-defined approach to quantum gravity, and after the first few steps in its direction it has been basically abandoned. Still, the initial motivations for going in its direction remain valid, in our opinion, and it is worth keeping them in mind. Therefore, we believe it is worthwhile to present this framework in some detail. Moreover, progress in loop quantum gravity allows to understand better what such a third quantization may mean and to clarify at least part of its construction. Most importantly, as we mentioned, group field theories can be understood as a way to realize this third quantization program by making it discrete and local, even if  at the cost of introducing a language that is farther away from that of General Relativity. The GFT formalism itself, despite recent developments, is still in its infancy and not well-known outside the community of researchers  working in loop quantum gravity and spin foam models, the area where it has found most applications, and even within this quantum gravity community, GFTs are still widely regarded as a purely technical tool, so that many of their conceptual aspects are not often stressed.  By direct formal comparison with continuum third quantization we will try to clarify here some of these conceptual issues: the ``level of quantization'' adopted, the consequent interpretation of the GFT classical action, the role of topology change in the framework, how the usual canonical quantum theory of gravity should be recovered, the difference between the ``global'' nature of the traditional quantization scheme (dynamics of a quantum universe) versus the more ``local'' nature of the GFT approach (quantizing building blocks of the same universe), and the implementation of spacetime symmetries into the GFT context.

At the more technical level,  our analysis of the third quantized framework, even if incomplete and confined to the level of a classical field theory on the space of connections, will be new and potentially useful for further developments. Among them, we have in mind both the study of the continuum limit of GFT models, which we will discuss in some detail, and the construction of simplified ``third quantized'' models, in contexts where they can be made amenable to actual calculations, for example in the context of quantum cosmology.  

Finally, apart from concrete results of our analysis, one more goal is to stimulate, motivate and suggest to some extent a possible direction for work on the continuum approximation of GFTs. We will sketch possible lines of research aimed at elucidating this issue, and speculate on how one can possibly derive one framework from the other, on how the continuum limit of GFTs can be related to third quantized continuum gravity, what such a continuum limit may look like, and on the emergence of classical and quantum General Relativity from it. In the end, the hope is to learn something useful about GFT from the comparison with a more formal framework, but one that is also closer to the language of General Relativity.

\subsection{Canonical quantization, loop quantum gravity and the idea of third quantization}

As we have mentioned above, the idea of third quantization of General Relativity arose from, and is closely connected to, the canonical quantization program and its modern formulation, loop quantum gravity.

 In canonical (Dirac) quantization of a classical theory, one takes the classical phase space, a symplectic manifold, and imposes the Poisson brackets between (a subalgebra of) classical observables (functions on phase space) as commutators of Hermitian operators. Then one chooses a polarization of the phase space, essentially a splitting into ``coordinates" and ``momenta," and defines a Hilbert space of wavefunctions depending on coordinates (or momenta) only. In the case of a constrained system, the (first class) constraints are imposed as operator equations determining a projection to the physical Hilbert space. Wheeler's geometrodyamics program \cite{geometro} was an early attempt to apply this procedure to General Relativity in 4 spacetime dimensions, in metric variables, on a manifold of given topology $\M=\Sigma\times\mathbb{R}$. The space of ``coordinates" one would like to use for quantization is {\it superspace}. This is usually defined as the space of Riemannian geometries on $\Sigma$, \ie metrics modulo spatial diffeomorphisms  (as we will see in section \ref{contsect}, one can also work with the larger space of metrics on $\Sigma$ and include the invariance under spatial diffeomorphisms as one of the constraints imposed by the equations of motion of the third quantized theory; this is indeed what happens in the 3d example we will discuss below).\footnote{There is of course no relation to the ``superspace" obtained by adding anticommuting coordinates to spacetime in supersymmetric theories.} Classically, the Hamiltonian determines the constraint surface in superspace and determines gauge motions in this surface, describing the evolution of a spatial hypersurface $\Sigma$ in ``time." Quantum-mechanically one is led to the infamous Wheeler-DeWitt equation \cite{dewitt},
\ben
\mathcal{H}(x)\Psi[h_{ij}]\equiv\left[\mathcal{G}^{ijkl}[h](x)\frac{\delta^2}{\delta h_{ij}(x)\delta h_{kl}(x)}-(R[h](x)-2\Lambda)\right]\Psi[h_{ij}]=0,
\label{wdw}
\een
which is the analogue of the classical Hamiltonian constraint generating reparametrization of the time coordinate. Here $\mathcal{G}^{ijkl}=\sqrt{h}h^{ij}h^{kl}+\ldots$ is the DeWitt supermetric and the wavefunctional $\Psi$ depends on the 3-dimensional metric $h_{ij}$ which encodes the geometry of $\Sigma$. 

(\ref{wdw}), of course, is mathematically ill-defined as an operator equation on a supposed ``Hilbert space" of wavefunctions $\Psi[h_{ij}]$, and suffers from severe interpretational problems. Geometrodynamics therefore never made much progress as a physical theory. Nevertheless, quite a lot is known about geometrical and topological properties of superspace itself and how it relates to (classical) geometrodynamics formulated on it; for a nice review see \cite{giulinisuper}.
The significance of superspace as an arena for quantum gravity is an important point to understand, so let us stress it again. While from the conceptual point of view {\it superspace} is to be understood as a kind of \lq\lq meta-space\rq\rq, or \lq\lq a space of spaces\rq\rq , in the sense that each of its ``points'', a 3-metric, represents a possible physical space, from the mathematical point of view it is a metric manifold in its own right, with a given fixed metric and a given topology. This is the key fact that makes possible to combine the background independence with respect to physical spacetime required by General Relativity with the use of the background dependent tools of (almost) ordinary quantum field theory in a third quantized field theory formalism of the type we will discuss below. This key aspect is shared also by group field theories and is one of the reasons for hoping they can lead to much further progress.

One may try to simplify matters by assuming a limited class (say homogeneous and isotropic) of possible metrics $h_{ij}$ and thereby picking out a certain, usually finite-dimensional, submanifold (``minisuperspace'') in superspace. While this leads to a well-defined quantum theory, it is not entirely clear how this simplified dynamical system is related to full quantum gravity.

When in the 1980s Ashtekar gave a reformulation of General Relativity in connection variables \cite{ashtekar} that showed more promise for quantization, the canonical quantization program experienced a revival in the form of loop quantum gravity (LQG) \cite{rovelli,thomas}. In the connection formulation superspace is replaced by the space of $\frak{g}$-connections on $\Sigma$, where $G$ is the gauge group of the theory ($G=\SU(2)$ in Ashtekar's formulation). This space has been studied mostly as the configuration space for Yang-Mills theory. One then takes the following steps towards quantization:
\begin{itemize}
\item Change variables in the classical phase space, going from connections to ($G$-valued) holonomies, exploiting the fact that a connection can be reconstructed if all of its holonomies along paths are known. The canonically conjugate variable to the connection, a triad (frame field), is replaced by its ($\frak{g}$-valued) fluxes through surfaces.
\item Compute Poisson brackets of the classical variables; in terms of the new variables on the phase space one finds commutativity among cylindrical functions of group variables (holonomies) and non-commutativity among Lie algebra variables (fluxes). In the simplest case, one considers a fixed graph $\Gamma$ embedded into $\Sigma$ and takes cylindrical functions $\Phi_f$ of holonomies along the edges of the graph (see section \ref{holonomy} on cylindrical functions) as well as fluxes $E^i_e$ through surfaces intersecting the graph only at a single edge $e$ as elementary variables. One then finds the algebraic structure, for each edge, of $T^*G\simeq G\times\frak{g}$, the cotangent bundle over $G$  \cite{noncomm},
\[
\{\Phi_f,\Phi_{f'}\}=0, \quad \{E^i_e,E^j_{e'}\}=\delta_{e,e'}{C^{ij}}_k E^k_e,
\]
where ${C^{ij}}_k$ are the structure constants of the Lie algebra.
\\One hence has two possible representations for wavefunctions: One where wavefunctions are functionals of the connection, and a non-commutative flux representation \cite{fluxrep}. A third representation arises when functionals of the connection are decomposed into irreducible representations of $G$. (These three representations also exist for GFT, as we show in section \ref{gftsect}.)
\item In the connection formulation, one defines a Hilbert space of functionals of (generalized) $\frak{g}$-connections by decomposing such functionals into sums of cylindrical functions which only depend on a finite number of holonomies, associated to a given graph, each: $\frak{H}\sim\bigoplus_{\Gamma}\frak{H}_{\Gamma}$ (again see section \ref{holonomy} for details). In this representation, one defines holonomy operators acting by multiplication and flux operators as left-invariant vector fields on $G$, hence implementing their non-commutativity.
\item Define the action of (in 4d LQG) diffeomorphism and Gauss constraints on cylindrical functions, and pass to a reduced Hilbert space of gauge invariant, (spatially) diffeomorphism invariant states.
\end{itemize}
While this procedure implements the kinematics of the theory rigorously, the issue of dynamics, \ie the right definition of the analogue of the Hamiltonian constraint (\ref{wdw}) on the kinematical Hilbert space and hence of the corresponding space of physical states, is to a large extent still an open issue. One now usually attempts to define it through a sum-over-histories \cite{carloreview}.

In connection variables, dimensionally reduced systems have been studied in the context of loop quantum cosmology (LQC) \cite{lqcreview}, where one mimics the above steps in a symmetry-reduced model.

What we have discussed so far is a ``first quantization", where the wavefunction is interpreted as giving probabilities for states of a single particle (here: a single hypersurface $\Sigma$). The form of (\ref{wdw}) suggests to draw an analogy to the case of a relativistic particle, where the mass-shell constraint $p^2+m^2=0$ leads to the wave equation
\ben
\left[g^{\mu\nu}(x)\frac{\partial}{\partial x^{\mu}}\frac{\partial}{\partial x^{\nu}}-m^2\right]\Psi(x)=0,
\label{kgor}
\een
the Klein-Gordon equation. The straightforward interpretation of $\Psi(x)$ as a single-particle wavefunction fails: In order to define on the kinematical Hilbert space a projection to the solutions of (\ref{kgor}) which has the correct composition properties and which on the solutions reduces to a positive definite inner product (for an overview of possible definitions for inner products and their composition laws, see \cite{halliort}; for analogous expressions in loop quantum cosmology see \cite{2pointcos}), one needs to define a splitting of the solutions to (\ref{kgor}) into positive- and negative-frequency solutions. This splitting relies on the existence of a timelike Killing vector $k$ and hence a conserved quantity $k\cdot p$ (``energy") on Minkowski space; as is well known, for generic metrics without isometries there is no unambiguous particle concept.

The usual interpretation of (\ref{kgor}) which is compatible with Lorentz invariance is to view $\Psi(x)$ as an {\em operator} field whose coefficients in an expansion into positive and negative frequency modes are interpreted as creation and annihilation operators. This leads to ``second quantization" where the particle concept is secondary; particles arise as excitations of the quantum field. 

The close analogy between (\ref{wdw}) and (\ref{kgor}) suggests that for a meaningful ``one-universe" concept in quantum geometrodynamics, one would need a conserved quantity, or at least one which is constant on all solutions of the constraints, on superspace. It was argued in \cite{kuchar} that no such quantity, and hence no consistent one-system interpretation of the quantum theory, exists. The conclusion is that one has to go to a many-geometries formalism in which ``universes," 3-manifolds of topology $\Sigma$, can be created and annihilated, and hence to a QFT on superspace. The analogy was pushed further, on a purely formal level, by Teitelboim \cite{teitelboim82} who gave analogues of the Feynman propagator, QFT perturbation theory etc. for such a tentative theory.

\

The general idea is to define a (scalar) field theory on superspace $\mathcal{S}$ for a given choice of
spatial manifold topology $\Sigma$, \eg the 3-sphere, essentially turning the wave function of the canonical (first quantized) theory into an operator $\Phi[^3h]$, whose dynamics is defined by an action with a kinetic term of the type:
\be
S_{{\rm free}}(\Phi)=\int_{\mathcal{S}}\mathcal{D} ^3h\,\Phi[^3h]\mathcal{H}\Phi[^3h] 
\ee
with $\mathcal{H}$ being the Wheeler-DeWitt differential operator of canonical
gravity (\ref{wdw}) here defining the free propagation of the theory. 
One thinks of a quantum field which is a functional $\Phi[^3h]$ of the 3-metric defined on $\Sigma$; the operator $\Phi[^3h]$  creates a 3-manifold of topology $\Sigma$ with metric $h$.

The quantum theory
would be \lq\lq defined\rq\rq\, by the partition function (``functional functional integral") \mbox{$Z=\int\frak{D}\Phi\, e^{-S(\Phi)}$}, in its perturbative expansion in \lq\lq Feynman diagrams\rq\rq. Adding interaction terms to the action, \eg a term cubic in $\Phi$, one would encounter diagrams corresponding to processes such as the following:

\begin{figure}[htp]
\centering
  \begin{picture}(267,73)
  \bezier{567}(0,7)(20,0)(40,7)\bezier{567}(0,7)(20,13)(40,7)
  \bezier{567}(60,7)(80,0)(100,7)\bezier{567}(60,7)(80,13)(100,7)
  \bezier{300}(40,7)(40,20)(50,27)\bezier{300}(60,7)(60,20)(50,27)
  \bezier{300}(0,7)(30,47)(30,67)\bezier{300}(100,7)(70,47)(70,67)
  \bezier{567}(30,67)(50,60)(70,67)\bezier{567}(30,67)(50,73)(70,67)
  \put(117,40){$+$}
  \bezier{567}(133,7)(153,0)(173,7)\bezier{567}(133,7)(153,13)(173,7)
  \bezier{567}(193,7)(213,0)(233,7)\bezier{567}(193,7)(213,13)(233,7)
  \bezier{300}(173,7)(173,20)(183,27)\bezier{300}(193,7)(193,20)(183,27)
  \bezier{300}(133,7)(163,47)(163,67)\bezier{300}(233,7)(203,47)(203,67)
  \bezier{567}(163,67)(183,60)(203,67)\bezier{567}(163,67)(183,73)(203,67)
  \bezier{567}(163,40)(183,33)(203,40)\bezier{567}(170,39)(183,47)(197,39)
  \put(250,40){$+\;\ldots$}
  \end{picture}
\end{figure}

Thus such a formalism also has the attractive aspect of incorporating topology change. In the simplified setting of homogeneous 3-spheres (\ie a third quantized minisuperspace model), it was explored by Giddings and Strominger in \cite{giddstrom} in the hope to find a dynamical mechanism determining the value of the cosmological constant, under the name ``third quantization."

The Feynman amplitudes will be given by the quantum gravity path integral (sum over geometries) for each spacetime topology (identified with a particular interaction process of universes), with the one for trivial topology representing a sort
of one particle propagator, thus a Green function for the
Wheeler-DeWitt equation. Other features of this (very) formal setting are: 1) the
full classical equations of motions for the scalar field on superspace will be a
non-linear extension of the Wheeler-DeWitt equation of canonical
gravity, due to the interaction term in the action, \ie the
inclusion of topology change; 2) the perturbative 3rd quantized vacuum
of the theory will be the \lq\lq no space\rq\rq\, state, and not any
state with a semiclassical interpretation in terms of a
smooth geometry, say a Minkowski state.

In the third quantization approach one has to deal with the quantization of a classical field theory defined on an infinite-dimensional manifold, clearly a hopeless task. In a connection formulation of GR where superspace is replaced by the space of connections, there is more hope of at least defining the classical theory, due to the work done in LQG. We will in fact show that the results of LQG allow at least to give a meaning to the free part of the third quantized action; notice in fact that the formal, mathematical complexity of the ``classical''\, third quantized action is the same as that of the canonical second quantized theory, \ie canonical quantum gravity. 

\

To make more progress, one will have to reduce the complexity of the system. One possibility would be to reduce to a symmetry-reduced sector of GR before quantization, obtaining a third quantized minisuperspace model, as done in metric variables in \cite{giddstrom} and for connection variables in \cite{gftc}. Another possibility, at least superficially quite different from a minisuperspace approximation, is the following: Instead of a continuous manifold $\Sigma$, consider a discrete structure such as a $D$-simplex where one is only interested in group elements characterizing the holonomies of the connection along the edges. The configuration space is then $G^3$ for a 2-simplex, for instance, and one can again define a standard field theory on a finite-dimensional space. Upon third quantization, such a theory, known as group field theory, is interpreted as describing many interacting ``building blocks" of space. GFTs are an attractive approach to resolving some of the technical and the interpretational problems of a field theory on the space of connections. Before getting to the GFT setting, we will show how the third quantization idea is implemented rigorously in an even simpler context, that of 2d quantum gravity, by means of matrix models.

We note that a third quantized theory describing the interaction of manifolds of given topology $\Sigma$ is also closely related to string field theory, where $\Sigma=S^1$ (and one has additional fields in addition to 2d gravity). Thus, this tentative framework would have the potential of relating different approaches to quantum gravity in different limits. In the following, we will investigate the connections to matrix models, group field theory and loop quantum gravity in particular. 

\sect{Matrix models: a success story}
\label{matrmod}
A simple context in which the idea of ``third quantization'' of gravity, a covariant formulation of the dynamics of geometry that also includes a dynamical topology in a field theory language, can be realized rigorously, is that of 2d Riemannian quantum gravity. The formalism that manages this is that of matrix models. The way this is achieved is the same ``go local, go discrete'' that characterizes group field theories. Indeed, the latter can be seen as a higher dimensional generalization for the same formalism (with some added structures).

Define a simple action for an $N\times N$ Hermitian matrix $M$, given by \bes S(M) &=& \frac{1}{2} \Tr M^2 \, -\, \frac{g}{\sqrt{N}}\, \Tr M^3 \, = \frac{1}{2}  M^i {}_{j}  M^j {}_{i} \, - \frac{g}{\sqrt{N}}  M^i {}_{j}  M^j {}_{k} M^k {}_i \, = \nonumber \\ &=&\, \frac{1}{2}  M^i {}_{j}  K_{j k l i} M^k {}_{l} \, - \frac{g}{\sqrt{N}}  M^i {}_{j}  M^m {}_{n} M^k {}_l \, V_{j m k n l i} \nonumber \\  && \text{ with} \hspace{1cm} K_{jkli} \,=\, \delta^j {}_k\, \delta^l {}_i \;\;\;\;\;\;\;\;   V_{j m k n l i}\,=\,  \delta^j {}_m \,\delta^n {}_k\, \delta^l {}_i \;\;\;\;\;\;\;\left( K^{-1} \right)_{j k l i} \, =\, K_{j k l i} .\ees  

The propagator and vertex term can be represented graphically as follows:

\vspace{0.5cm}
%\begin{figure}[here]
\hspace{3cm}
\begin{minipage}[c]{4cm}
$ \left( K^{-1}\right)_{jkli}$ \\   {\vspace{0.5cm}} \\  $ V_{j m k n l i} $
\end{minipage}
\hspace{0.2cm}
\begin{minipage}[l]{5cm}
\includegraphics[width=3.5cm, height=2.5cm]{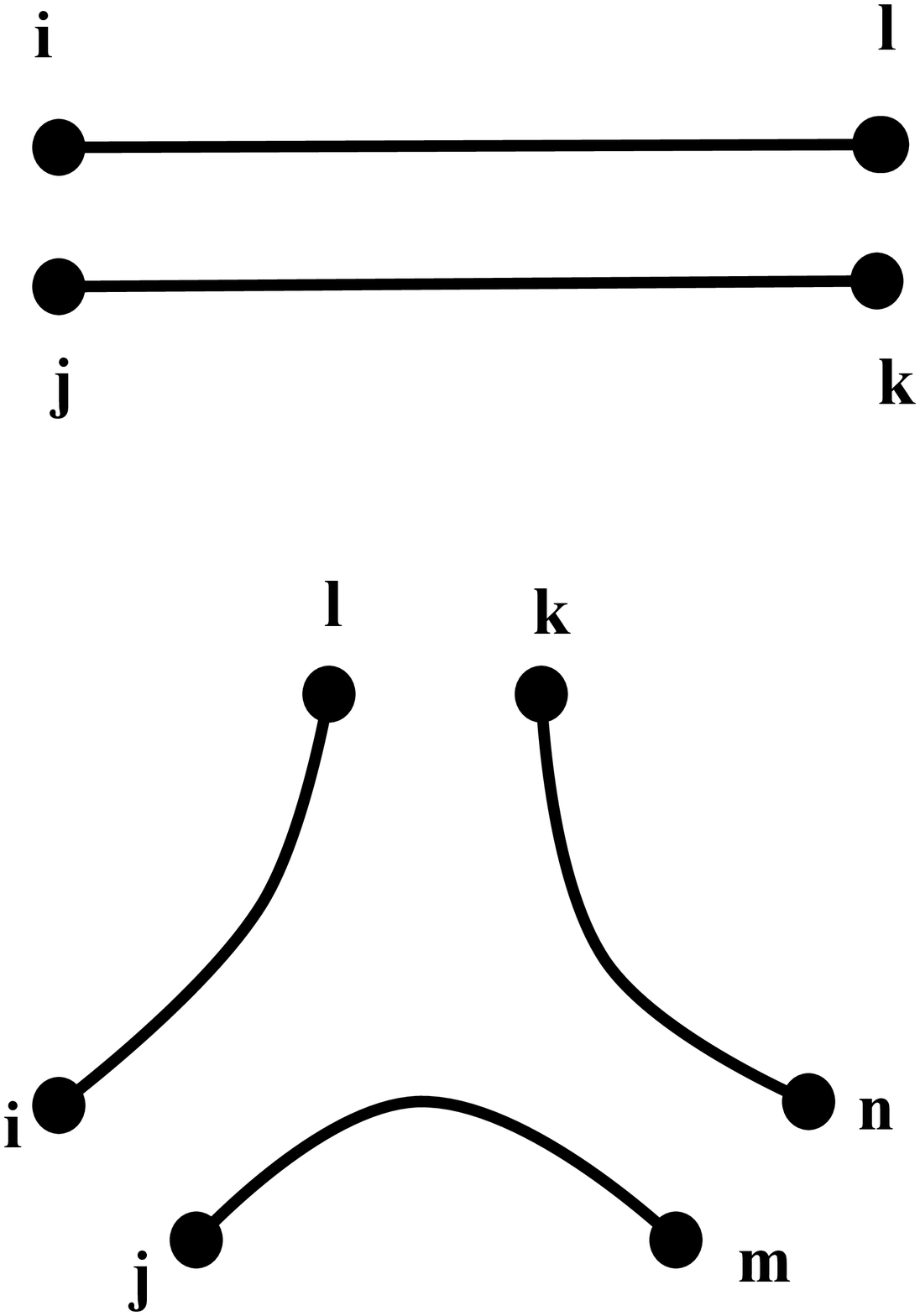}
\end{minipage}
%\caption{Building blocks of Feynman diagrams for a matrix model}
%\end{figure}
\vspace{0.5cm}

The composition of such building blocks is given by the tracing of indices $i,j,k$ in the kinetic and vertex term and represents identification of the points labeled by the same indices in the corresponding graphical representation. Feynman diagrams are then made of: (double) lines of propagation (made of two strands), non-local ``vertices'' of interaction (providing a re-routing of strands), faces (closed loops of strands) obtained after index contractions. This combinatorics of indices (and thus of matrices)  can be given a simplicial representation as well, by interpreting a matrix as representing an edge in a 2-dimensional simplicial complex:

\begin{figure}[here]
 \includegraphics[width=9cm, height=3cm]{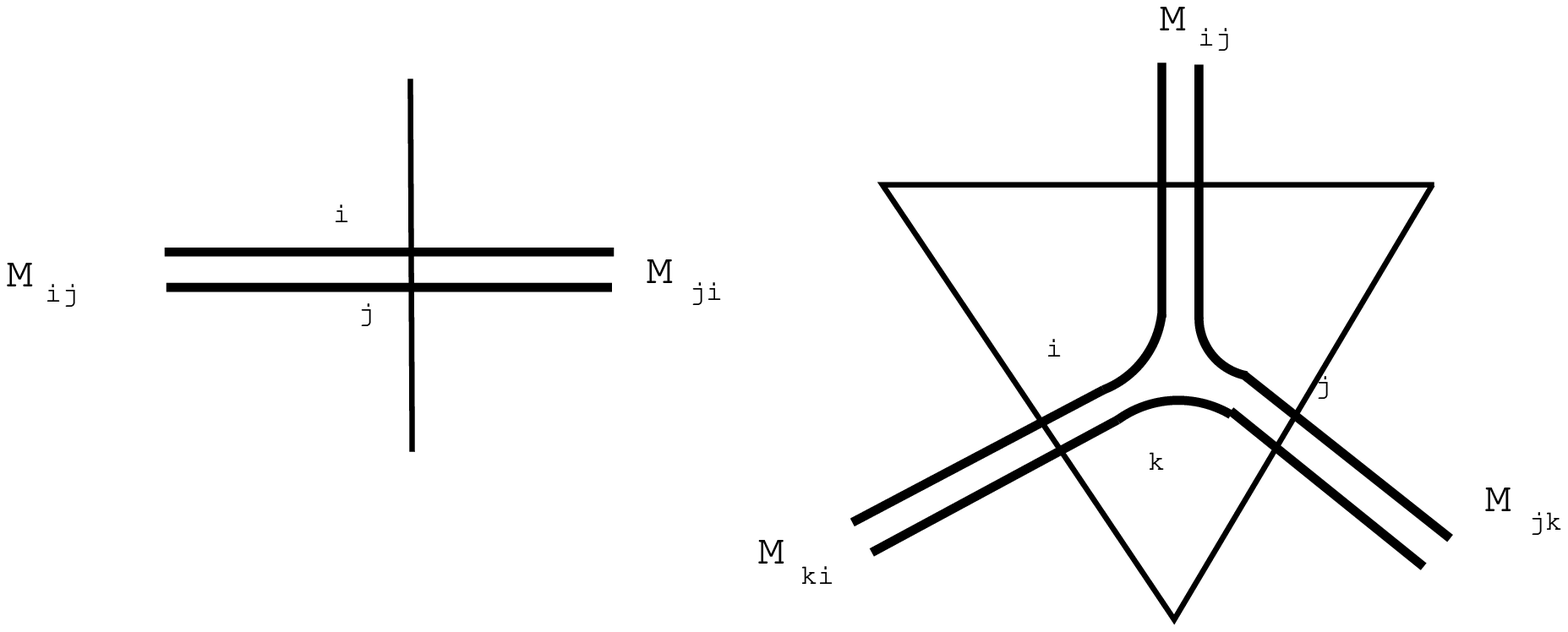}
\end{figure}

Therefore the diagrams used in evaluating the partition function $Z = \int dM_{ij} \, e^{- S(M)}$ correspond to such complexes of arbitrary topology, obtained by arbitrary gluing of edges to form triangles (in the interaction vertex) and of these triangles to one another along common edges (as dictated by the propagator). A discrete 2d spacetime emerges as a virtual construction, encoding the possible interaction processes of fundamentally discrete quanta of space.

The partition function, expressed in terms of Feynman amplitudes for this model, is then
$$ Z\, =\,  \sum_\Gamma\, g^{V_\Gamma} N^{F_\Gamma - \frac{1}{2} V_\Gamma}=\sum_\Gamma\, g^{V_\Gamma}\, N^{\chi}, $$
where $V_\Gamma$ is the number of vertices and $F_\Gamma$ the number of faces of the Feynman graph $\Gamma$, $\chi$ the Euler characteristic of the simplicial complex, and $N$, again, the dimension of the matrices.

We now ask what is the relation of this with (simplicial) gravity. Each Feynman amplitude will be associated to simplicial path integrals for gravity discretized on the associated simplicial complex $\Delta$. Continuum (Riemannian) 2d GR with cosmological constant $\Lambda$ on a 2d manifold $S$ has the action $ \frac{1}{G}\int_{S} d^2 x \,\, \sqrt{g}\, \left( - R (g) \, + \, \Lambda \right)\, =\, -\, \frac{4\pi}{G} \, \chi\, +\, \frac{\Lambda}{G}\, A_S = - \, \frac{4\pi}{G} \, \chi\, +\, \frac{\Lambda a}{G} \, t$, where $A_S$ is the area of $S$ and in the last equality we think of $S$ as discretized into $t$ equilateral triangles of area $a$. Defining $g=e^{-\frac{\Lambda a}{G}}$ and $ N = e^{+\, \frac{4 \pi}{G}} $, we can identify ($t_\Delta = V_\Gamma$ since each vertex is dual to a triangle):
$$
Z\,=\, \sum_\Gamma\, g^{V_\Gamma}\, N^{\chi}\, =\, \sum_{\Delta}\, e^{+\frac{4\pi}{G}\chi(\Delta) \, -\, \frac{a\Lambda}{G} \, t_\Delta}.
$$
In other words, we obtain a (trivial) sum over histories of discrete GR on given 2d complex, whose triviality is due to the fact that the only geometric variable associated to each surface is  its area, the rest being only a function of topology. In addition to this sum over geometries, from our matrix model we obtain a sum over all possible 2d complexes of all topologies. In other words, the matrix model defines a discrete third quantization of GR in 2d!

Can we control this sum over triangulations and over topologies? Yes, because \cite{mm1,mm2} the sum is governed only by topological parameters, namely the genus $h$,
$$
Z\,=\, \sum_\Delta\, g^{t_\Delta}\, N^{\chi(\Delta)}\,=\, \sum_\Delta\, g^{t_\Delta}\, N^{2 - 2 h}  =\sum_h N^{2- 2 h }\, Z_h(g)\,=\,N^2 \, Z_0(g)\,+\, Z_1(g)\,+ N^{-2}\, Z_2(g) + \ldots
$$

It is then apparent that, in the limit $N\rightarrow\infty$, only spherical simplicial complexes (those of trivial topology, with genus 0) contribute significantly to the sum.

The second question is whether, as $N\rightarrow\infty$, one can also define a continuum limit and match the results of the continuum 2d gravity path integral. In order to study the continuum limit we expand $Z_0(g)$ in powers of $g$,
$$ Z_0(g) = \sum_{t} t^{\gamma -3} \left(\frac{g}{g_c}\right)^{t} \simeq_{t\rightarrow\infty} \left( g - g_c\right)^{2 - \gamma},$$
where $\gamma$ is a critical exponent, so that, in the limit of large number of triangles, and for the coupling constant approaching the critical value $g \rightarrow g_c$ ($\gamma >2$), the partition function diverges. 
This is a signal of a phase transition. In order to identify his phase transition as continuum limit we compute the expectation value for the area of the surface:
$ \la A \ra =\, a\, \la t_\Delta \ra = a \frac{\partial}{\partial g} \ln{ Z_0(g)} \, \simeq\, \frac{a}{g- g_c}$, for large $t$. Thus we can send the area of each triangle to zero, $a \rightarrow 0$, and the number of triangles to infinity: $t \rightarrow \infty$ (continuum limit), while tuning at the same time the coupling constant to $g_c$, to get finite continuum macroscopic areas.
This defines a continuum limit of the matrix model, which reproduces \cite{mm1,mm2} the results obtained from a continuum 2d gravity path integral (when this can be computed); otherwise the matrix model can be taken to {\em define} the continuum path integral.

Let us now ask whether the 3rd quantized framework we have (in this 2d case) also allows to understand and compute the contribution from non-trivial topologies in a continuum limit, thus defining a {\it continuum} 3rd quantization of 2d gravity.
The key technique is the so-called  double-scaling limit \cite{mm1,mm2}. One can first of all show that the contribution of each given topology of genus $h$ to the partition function is 
$$ Z_h(g) \simeq \sum_t t^{\frac{(\beta - 2) \chi}{2} - 1}  \left( \frac{g}{g_c}\right)^t \simeq \, f_h \, \left( g - g_c \right)^{\frac{(2 -\beta) \chi}{2}},$$ where the last approximation holds in the limit of many triangles (necessary for any continuum limit, corresponding to the thermodynamic limit), and where $\beta$ is a constant that can be computed.
One can now define $\kappa^{-1} = N \left( g - g_c \right)^{\frac{(2 -\beta)}{2}}$, so that in the thermodynamic limit we have an expression $Z \, \simeq\, \sum_h\, \kappa^{2 h - 2} f_h$. We can then take the combined limits $N\rightarrow\infty$ and $g \rightarrow g_c$, while holding $\kappa$ fixed. The result of this double limit is indeed a continuum theory to which all topologies contribute. 

\sect{Group field theory: a sketchy introduction}
\label{gftsect}
One can construct combinatorial generalizations of matrix models to higher tensor models whose Feynman diagrams will be higher simplicial complexes. It turns out however that such tensor models do not possess any of the nice scaling limits (large-$N$ limit,  continuum and double scaling limit) that allow to control the sum over topologies in matrix models and to relate their quantum amplitudes to those of continuum gravity. These difficulties suggest the need to generalize matrix and tensor models by adding degrees of freedom, \ie defining corresponding field theories. 
In the process, the indices of the tensor models will be replaced by continuous variables (valued in some domain space), while maintaining their combinatorial pairing in the action. For the definition of good models for quantum gravity, requiring a careful choice of domain space and classical action (kinetic/vertex functions), the input from other approaches to quantum gravity (loop quantum gravity, simplicial quantum gravity, non-commutative geometry) is crucial. 

\

We now describe in some more detail the GFT formalism in the case of 3d Riemannian gravity. We hope that the general features of the formalism will become clear after having considered this specific example.

\
Consider a triangle in $\mathbb{R}^3$. We consider its (2nd quantized) kinematics to be encoded in a (real) GFT field $\varphi$, understood as a function on the space of possible geometries for the triangle. We parametrize the possible geometries for the triangle in terms of three $\su(2)$ Lie algebra elements attached to its three edges, interpreted as fundamental triad variables obtained by discretization of continuum triad fields along the edges of the triangle: 
$$
\varphi \,:\, (x_1,x_2,x_3) \in \su(2)^3 \longrightarrow \varphi(x_1,x_2,x_3) \in \mathbb{R} 
$$ 

Using the non-commutative group Fourier transform \cite{PR3,NCFourier, danielearistide}, based on the plane waves  $e_{g}(x) \!=\! e^{i \vec{p}_g\cdot \vec{x}}$ as functions on $\g \sim \R^n$, depending on a choice of coordinates  $\vec{p}_g = \Tr g \vec{\tau}$ on the group manifold, the GFT field can be recast as a function of $\SU(2)$ group elements: 
\[
\varphi(x_1, x_2, x_3) =\int [\extd g]^3\, \vphi(g_1, g_2, g_3) \, e_{g_1}(x_1) e_{g_2}(x_2) e_{g_3}(x_3) 
\]
so that the GFT field can also be seen as a function of three group elements, thought of as parallel transports of the gravity connection along fundamental links dual to the edges of the triangle represented by $\varphi$, which intersect the triangle only at a single point.

In order to define a geometric triangle, the vectors (Lie algebra elements) associated to its edges have to ``close'' to form a triangle, \ie they have to sum to zero. We thus impose the constraint ($\star$ is a non-commutative star product reflecting the non-commutativity of the group multiplication law in Lie algebra variables)
\ben
\vphi = C \star \vphi, \quad C(x_1, x_2, x_3) = \delta_0(x_1 \!+\!x_2\!+\!x_3) := \int \extd g \, e_g(x_1+x_2+x_3).
\label{closure}
\een

In terms of the dual field $\varphi(g_1,g_2,g_3)$, the closure constraint (\ref{closure}) implies invariance under the diagonal (left) action of the group $\SU(2)$ on the three group arguments, imposed by projection $P$:
\ben
\varphi(g_1,g_2,g_3) = P \varphi(g_1,g_2,g_3) = \int_{\SU(2)} \extd h\, \phi( h g_1, h g_2, h g_3) \label{invariance}
\een

Because of this gauge invariance, which is in fact imposed in the same way as the Gauss constraint is imposed on cylindrical functions in LQG, the field can be best depicted graphically as a 3-valent vertex (on which gauge transformations can act) with three links, dual to the three edges of the closed triangle (Fig. \ref{gftfig}). This object, both mathematically and graphically, will be the GFT building block of our quantum space.

One obtains a third representation of the GFT field by means of Peter-Weyl decomposition into irreducible representations, in the same way as one obtains the spin network expansion of generic cylindrical functions in LQG (cf. section \ref{holonomy}).
The gauge invariant field decomposes in $\SU(2)$ representations as:

\begin{equation}
\varphi(g_1,g_2,g_3) = \sum_{j_1,j_2,j_3} \, \varphi^{j_1j_2j_3}_{m_1m_2m_3}\, D^{j_1}_{m_1n_1}(g_1)D^{j_2}_{m_2n_2}(g_2)D^{j_3}_{m_3n_3}(g_3)\,C^{j_1 j_2 j_3}_{n_1 n_2 n_3}
\label{peterweyl}
\end{equation}
where $ C^{j_1 j_2 j_3}_{n_1 n_2 n_3}$ is the Wigner invariant 3-tensor, the 3j-symbol. Graphically, one can think of the GFT field in any of the three representations (Lie algebra, group, representation), as appropriate:

\begin{figure}[here]
\includegraphics[width=13cm, height=3cm]{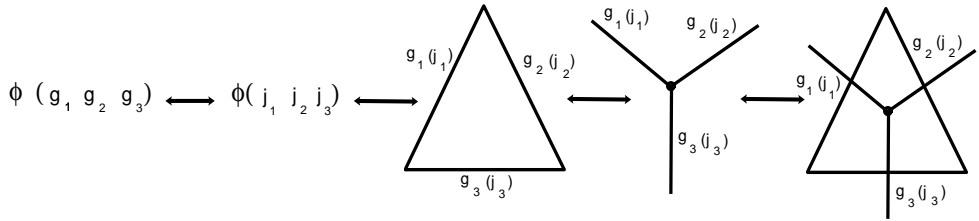}
\caption{The GFT field $\varphi$ has different representations which describe the same geometric structure in equivalent ways.}
\label{gftfig}
\end{figure}

Multiple fields can be convoluted (in the group or Lie algebra picture) or traced (in the representation picture) with respect to some common argument. This represents the gluing of multiple triangles along common edges, and thus the formation of more complex simplicial structures, or, dually, of more complicated graphs:
\begin{figure}[here]
\includegraphics[width=13cm, height=3cm]{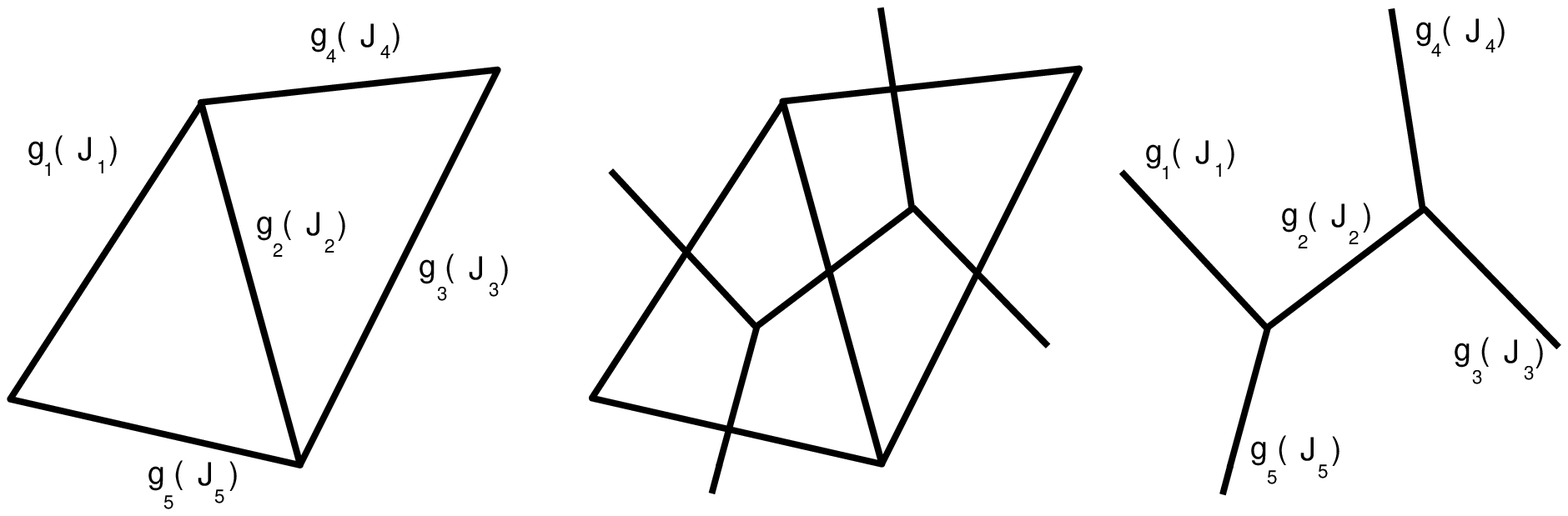}
\end{figure}

The corresponding field configurations represent thus extended chunks of quantum space, or many-GFT-particle states. A generic polynomial GFT observable would be associated with a particular quantum space.
This includes, of course, any open configuration, in which the arguments of the involved GFT fields are not all convoluted or contracted, representing a quantum space (not necessarily connected) with boundary. 

\

We now define a classical dynamics for the introduced GFT field. The prescription for the interaction term, as in tensor models, is simple: four geometric triangles should be glued to one another, along common edges, to form a 3-dimensional geometric tetrahedron.  The kinetic term should encode the gluing of two tetrahedra along common triangles, by identification of their edge variables. There is no other dynamical requirement at this stage. Defining the shorthand $\varphi_{123}:=\varphi(x_1,x_2,x_3)$, we are hence motivated to define the following structure of the action,
%One can also show that the $\star$-product of dual fields corresponds to the convolution product of fields: $\int\widehat{\vphi \circ \psi} =  \int\widehat{\vphi} \star %%\widehat{\psi}$. 
\ben
S =\! \frac{1}{2}\int [\extd x]^3 \,\varphi_{123} \star \varphi_{123} -  
\frac{\lambda}{4!} \! \int [\extd x]^6 \,\varphi_{123} \star \varphi_{345} \star \varphi_{526} \star \varphi_{641}
\label{gftaction}
\een
where it is understood that $\star$-products relate repeated indices as $\phi_i \star \phi_i \! :=\! (\phi \star \phi_{-})(x_i)$, with $\phi_{-}(x) \! = \! \phi(- x)$.

We would like to point out at this point that, as in the case of tensor and matrix models, combinatorial generalizations can be considered, since there is no a priori restriction on how many arguments a GFT field can have. Once closure constraint (\ref{closure}) or gauge invariance (\ref{invariance}) has been imposed on such a generalized field with $n$ arguments, it can be taken to represent a general $n$-gon (dual to an $n$-valent vertex) and glued to other fields in order to give a polygonized quantum space in the same way as we have outlined for triangles. Indeed such a generalization will be suggested to us by the considerations in section \ref{contsect} about a possible third quantization of continuum gravity. Notice also that there is no difficulty in dealing with a combinatorial generalization of the action in which, for given building block $\varphi(x_i)$, one adds other interaction terms corresponding to the gluing of triangles (or polygons) to form general polyhedra or even more pathological configurations (\eg with multiple identifications among triangles). The only restriction may come from the symmetries of the action.

The projection (\ref{invariance}) takes into account parallel transport between different frames. Combining the projection (\ref{invariance}) with the kinetic and interaction terms in (\ref{gftaction}), we can identify a propagator and a vertex:
\be \label{prop&vertex}
\mathcal{K}(x_i, y_i) = \int \extd h_t \, \prod_{i=1}^3 (\delta_{- x_i} \star \, e_{h_t} )(y_i), 
\quad 
\mathcal{V}(x_i,y_i) = \int \prod_t \extd h_t \, \prod_{i=1}^6 ( \delta_{- x_i} \star\, e_{h_{tt'}})(y_i),
\ee
with $h_{tt'} := h_{t\tau} h_{\tau t'}$, where we have used `$t$' for triangle and `$\tau$' for tetrahedron. The group variables $h_{t}$ and $h_{t\tau}$ arise from (\ref{invariance}), and should be interpreted as parallel transports through the triangle $t$ for the former, and from the center of the tetrahedron $\tau$ to triangle $t$ for the latter. 
We may represent the propagator and vertex graphically as
%\begin{figure}[here]
%\includegraphics[width=6.5cm, height=5cm]{TensorPropVertex.eps}
%\end{figure}
\be \label{Feynmanrules}
\begin{array}{c}
% ---- propagator
\begin{tikzpicture}[scale=1.2]%[scale=2.4, rotate=90]
\draw (-0.3,-0.2) rectangle (0.3, 0.2);
\path (0, 0) node {\small $t$};
% in strands
\draw (-0.2, -1) -- (-0.2,-0.2);
\draw (0,-1) -- (0, -0.2);
\draw (0.2, -1) -- (0.2, -0.2);
\path (-0.2, -1.2) node {\tiny $x_1$};
\path (0, -1.2) node  {\tiny $x_2$};
\path (0.2, -1.2) node {\tiny $x_3$};
\path (0, -1.5) node  {\tiny $ $};

%\draw (0,0) rectangle (0.30, 0.40);
%\path (0.15, 0.2) node {\mbox{\small t}};
% in strands
%\draw (-0.4,0.05) -- (0,0.05);
%\draw (-0.4,0.20) -- (0,0.20); 
%\draw ( -0.4,0.35)-- (0,0.35);
%\path (-0.5, 0.35) node {\tiny $x_1$};
%\path (-0.5,0.20) node {\tiny $x_2$};
%\path (-0.5,0.05) node {\tiny $x_3$};
% out strands
\draw (-0.2, 0.2) -- (-0.2, 1);
\draw (0, 0.2) -- (0, 1);
\draw (0.2, 0.2) -- (0.2, 1);
\path (0.2, 1.2) node {\tiny $y_3$};
\path (0, 1.2) node  {\tiny $y_2$};
\path (-0.2, 1.2) node {\tiny $y_1$};
\path (0, 1.5) node  {\tiny $ $};

%\draw (0.30, 0.05) --(0.70, 0.05);
%\draw (0.30,0.20) --(0.70,0.20);
%\draw (0.30,0.35) --(0.70,0.35);
%\path (0.80,0.35) node {\tiny $y_1$};
%\path (0.80,0.20) node {\tiny $y_2$};
%\path (0.80,0.05) node {\tiny $y_3$};
\end{tikzpicture}
\hspace{1cm} 
% ---- vertex
\begin{tikzpicture}[scale=1.2]
\draw (-1, 0) -- (1,0);
\draw (0,-1) -- (0, -0.1);
\draw (0, 0.2) -- (0, 1);
\draw (-1, 0.2) -- (-0.2, 0.2);
\draw (0.2, 0.2) -- (1, 0.2);
\draw (-1, -0.2) -- (-0.2, -0.2);
\draw (0.2,- 0.2) -- (1, -0.2);
\draw (-0.2, -1) -- (-0.2,-0.2);
\draw (-0.2, 0.2) -- (-0.2, 1);
\draw (0.2, -1) -- (0.2, -0.2);
\draw (0.2, 0.2) -- (0.2, 1); 
\path (0, 0.1) node {\small $\tau$};
\path (-1.2, 0.2) node {\tiny $x_1$};
\path (-1.2, 0) node {\tiny $x_2$};
\path (-1.2, -0.2) node {\tiny $x_3$};
\path (-0.2, -1.2) node {\tiny $y_3$};
\path (0, -1.2) node  {\tiny $x_4$};
\path (0.2, -1.2) node {\tiny $x_5$};
\path (1.2, -0.2) node {\tiny $y_5$};
\path (1.2, 0) node {\tiny $y_2$};
\path (1.2, 0.2) node {\tiny $x_6$};
\path (0.2, 1.2) node {\tiny $y_6$};
\path (0, 1.2) node  {\tiny $y_4$};
\path (-0.2, 1.2) node {\tiny $y_1$};
\path (-1.5, 0) node {\tiny $t_a$};
\path (0, -1.4) node  {\tiny $t_b$};
\path (1.5, 0) node {\tiny $t_c$};
\path (0, 1.4) node  {\tiny $t_d$};
\end{tikzpicture}
\end{array}
\ee

The integrands in (\ref{prop&vertex}) factorize into a product of functions associated to strands (one for each field argument), with a clear geometrical meaning: the pair of variables $(x_i,y_i)$ associated to the same edge $i$ corresponds to the edge vectors seen from the frames associated to the two triangles $t, t'$ sharing it. The vertex functions state that the two variables are identified, up to parallel transport $h_{tt'}$, and up to a sign labeling the two opposite edge orientations inherited by the  triangles $t, t'$. The propagator encodes a similar gluing condition, allowing for the possibility of a further mismatch between the reference frames associated to the same triangle in two different tetrahedra.
In this graphical representation of the two terms in (\ref{gftaction}), the geometric content of the action is particularly transparent.

\

Using the group Fourier transform we obtain the form in which the model was originally defined by Boulatov \cite{boulatov},
\[ S[\phi]=\frac{1}{2}\int[\extd g]^3
\varphi(g_1,g_2,g_3)\varphi(g_3,g_2,g_1) \;-\;\frac{\lambda}{4!}\int [\extd g]^6
\varphi(g_{12},g_{13},g_{14})\varphi(g_{14},g_{24},g_{34})\varphi(g_{34},g_{13},g_{23})\varphi(g_{23},g_{24},g_{12}) . \]
In this group representation, the kinetic and vertex functions are
\begin{equation}
\mathcal{K}(g_i, \tilde{g}_i) = \int \extd h  \,\prod_{k=1}^{3}\,\delta(g_k h
\tilde{g}_k^{-1}), \hspace{1cm} \mathcal{V}(g_{ij},g_{ji}) =
\int \prod_{i=1}^4 \extd h_i \,\prod_{i<j}^{} \delta(g_{ij} h_i
h_j^{-1} g_{ji}^{-1}).
\label{kinpropgroup}
\end{equation}

Also in these set of variables, the geometric content of the model can be read out rather easily: the six delta functions in the vertex term encode the flatness of each \lq\lq wedge,\rq\rq\, \ie of the portion of each dual face inside a single tetrahedron. This flatness is characteristic of the piecewise-flat context  in which the GFT models are best understood.

Using the expansion of the field (\ref{peterweyl}), performing the appropriate contractions and using standard properties of the $3j$-symbols from $\SU(2)$ recoupling theory, the action becomes:

\ben
S(\varphi) \, =\, \frac{1}{2} \sum_{\{ j\}, \{ m\} }\, \varphi^{j_1 j_2 j_3}_{m_1 m_2 m_3}\,\varphi^{j_3 j_2 j_1}_{m_3 m_2 m_1}\;-\;\frac{\lambda}{4!}\,\sum_{\{ j\}, \{ m\} }\, \varphi^{j_1 j_2 j_3}_{m_1 m_2 m_3}\,\varphi^{j_3 j_4 j_5}_{m_3 m_4 m_5}\,\varphi^{j_5 j_2 j_6}_{m_5 m_2 m_6}\,\varphi^{j_6 j_4 j_1}_{m_6 m_4 m_1}\; \left\{
\begin{array}{ccc} 
j_1 &j_2 &j_3
\\ j_4 &j_5 &j_6 
\end{array}\right\}
\label{propvertJ}
\een

from which one can read out the kinetic and vertex terms.

\

The classical equations of motion for this model are, in group space:
$$
\int \extd{h}\,\phi(g_1h,g_2 h,g_3 h) \,-
\,\frac{\lambda}{3!}\,\int \prod_{i=1}^{3}  \extd h_{i}\int
\prod_{j=4}^{6} \extd g_j\, \phi(g_3 h_1,g_4 h_1,g_5 h_1)
\phi(g_5 h_2,g_6 h_2,g_2 h_2)\phi(g_6 h_3,
g_4 h_3, g_1h_3)\,=\,0
$$
but can of course be written both in Lie algebra space, where they also look like complicated integral equations, as well as in the form of purely algebraic equations in representation space.

These equations define the classical dynamics of the field theory; they would allow the identification of classical background configurations, non-trivial GFT phases, etc. From the point of view of quantum gravity, considering the geometric interpretation of the GFT as a discrete ``3rd quantization'' of gravity, the classical GFT equations encode fully the quantum dynamics of the underlying (simplicial) canonical quantum gravity theory, \ie the one in which quantum gravity wave functions are constructed, or equivalently the quantum dynamics of first quantized spin networks, thus implementing the Hamiltonian and diffeomorphism constraints of canonical GR. This should become clear once we have presented the formal third quantization of gravity in connection variables.

\

The quantum dynamics is defined by the perturbative expansion of the partition function in Feynman diagrams, which are viewed as dual to 3-dimensional simplicial complexes (in (\ref{Feynmanrules}), the propagator represents a triangle and the vertex a tetrahedron in such a simplicial complex):

$$ Z\,=\,\int
\mathcal{D}\varphi\,e^{-S[\varphi]}\,=\,\sum_{\Gamma}\,\frac{\lambda^N}{{\rm sym}[\Gamma]}\,Z(\Gamma),
$$
where $N$ is the number of interaction vertices in the Feynman graph
$\Gamma$, ${\rm sym}[\Gamma]$ is a symmetry factor for the graph and
$Z(\Gamma)$ the corresponding Feynman amplitude.

In Lie algebra variables, the Feynman amplitudes for a generic Feynman diagram take the form \cite{danielearistide}
\be \label{bf} Z(\Gamma) =  \int \prod_p \extd h_p \prod_f \extd x_f  \, e^{i \sum_f \Tr \, x_f H_f}, \ee
where we denote closed loops in the Feynman graph by $f$, since they bound faces of the 2-complex formed by the Feynman graph; such faces are dual to edges in the simplicial complex, and they are associated a Lie algebra variable $x_f$ playing the role of discretized triad. We also integrate over a group element $h_p$ for each propagator $p$ in the Feynman graph (corresponding to a triangle in the dual simplicial complex) interpreted as a discrete connection. Now the central result is that (\ref{bf}) is the usual expression for the simplicial path integral of 3d gravity in 1st order form. Consider the 3d gravity first order action in the continuum:
$$ S(e,\omega)\,=\,\int_{\mathcal{M}} \Tr \left( e\wedge F(\omega)\right) , $$  
with triad 1-form $e^i(x)$ (seen as valued in $\su(2)$) and a connection 1-form $\omega^{j}(x)$ also valued in $\su(2)$, with curvature 2-form $F(\omega)$. 
Introducing the simplicial complex $\Delta$ and its topological dual cellular complex $\Gamma$, we can discretize the triad in terms of Lie algebra elements associated to the edges of the simplicial complex as $x_e = \int_e e(x) = x^i \tau_i \in \su(2)$, and the connection in terms of elementary parallel transports along links of the dual complex $\Gamma$, dual to triangles of $\Delta$, as $h_{L} = e^{\int_{L}\omega}\in \SU(2)$. The discrete curvature is then given by the holonomy around the face $f$ in $\Gamma$ dual to an edge $e$ of $\Delta$, obtained as ordered product of group elements $h_L$ associated to its boundary links: $H_{f} = \prod_{L\in \partial f} h_{L}\in \SU(2)$, and a discrete counterpart of the continuum action will be given by the action $\sum_f \Tr \, x_f H_f$ appearing in (\ref{bf}).

The GFT model we have introduced succeeds in at least one of the points where the simpler tensor models failed: in defining amplitudes for its Feynman diagrams (identified with discrete spacetimes), arising in a perturbative expansion around the ``no-space state,'' that correctly encode classical and quantum simplicial geometry and that can be nicely related to a simplicial action for gravity. 

\

The Feynman amplitudes can also be computed in the other representations we have at our disposal. In the group picture the overall amplitude is:
\begin{equation}
Z(\Gamma)\,=\, \prod_{L\in \Gamma} \int \,\extd h_L
\,\,\prod_{f}\,\delta (\prod_{L\in\partial f} h_L )\, \label{amplgroup},
\end{equation}
which makes the flatness constraint explicit. Similarly, the expression of $Z(\Gamma)$ in terms of group representations (quantum numbers of geometry) can be computed starting from the action (\ref{propvertJ}), or by Peter-Weyl decomposition of (\ref{amplgroup}). The result is an assignment of an irreducible $\SU(2)$ representation $j_f$ to each face of $\Gamma$, and of a group intertwiner to each link of the complex, a {\em spin foam} \cite{review,alex,thesis}, whose amplitude reads:
$$ Z(\Gamma)=\left(\prod_{f}\,\sum_{j_{f}}\right)\,\prod_{f}(2j_{f} +1)\,\prod_{v}\, \left\{ \begin{array}{ccc}
j_1 &j_2 &j_3
\\ j_4 &j_5 &j_6
\end{array}\right\}. $$

This is the well-known Ponzano-Regge spin foam model \cite{ponzanoregge} for 3d Riemannian quantum gravity. The correspondence between spin foam models and GFT Feynman amplitudes exhibited in this example is generic.

\

All the above constructions can also be generalized to the computation of GFT observables, in particular $n$-point functions, which translate (in their perturbative expansion) into the calculation of Feynman amplitudes for diagrams/simplicial complexes of arbitrary topology and with boundaries. These in turn take again the form of simplicial gravity path integrals on the corresponding topology, with appropriate boundary terms.

Once more, we have a discrete realization of the third quantization idea. We refer to the literature (for example the recent, more extended introduction to the GFT formalism in \cite{ProcCapeTown}) for more details on the GFT framework, and for an account of recent results, in particular for the definition of GFT models of 4d quantum gravity, for the identification of (discrete) diffeomorphism symmetry at the GFT level, and for the proof that a generalization of the large-$N$ limit of matrix models  holds true also in (some) GFT models leading again to the suppression of topologies different from the simplest ones (thus solving another main problem of naive tensor models), as well as for more work on the topological properties of the GFT Feynman expansion.

\

In order to appreciate this fully, let us discuss in more detail what a third quantization of gravity may mean in terms of continuum variables.

\sect{Continuum third quantization: Heuristics}
\label{contsect}

Going back to the idea of a third quantization framework for continuum gravity, we would like to define a field theory on the configuration space of General Relativity that reproduces the constraints of canonical General Relativity through its equations of motion. Since such constraints have to be satisfied at each point in the spatial hypersurface $\Sigma$ (where spacetime has topology $\Sigma\times \bR$), we should define an action which reproduces equations of motion of the form ($\gamma_\alpha$ are the dynamical variables of GR, metric or connection)
\ben
\mathcal{C}_i\left[\gamma_\alpha,\frac{\delta}{\delta \gamma_\alpha};x\right)\Phi[\gamma_\alpha]=0.
\label{constreq}
\een
Here $\mathcal{C}_i$ are constraints defining GR as a fully constrained system and we have introduced a field which is classically a functional of the variables $\gamma_\alpha$, \ie a canonical quantum gravity wavefunction. Defining the naive action
\ben
S=\int \mathcal{D}\gamma_\alpha\; \sum_i\Phi[\gamma_\alpha]\mathcal{C}_i\left[\gamma_\alpha,\frac{\delta}{\delta \gamma_\alpha}\right]\Phi[\gamma_\alpha]
\een
can only reproduce constraints $\mathcal{C}_i=0$ which are necessarily independent of the point on $\Sigma$. This was noted by \cite{giddstrom}, who took $\mathcal{C}$ to be the Hamiltonian constraint integrated over $\Sigma=S^3$ while assuming a homogeneous and isotropic geometry on $\Sigma$.

To get the right number of constraints, one has to increase the number of degrees of freedom of $\Phi$. One possibility, for only a single constraint per point $\mathcal{C}$, would be to add an explicit dependence of $\Phi$ on points in $\Sigma$:
\ben
S=\int \mathcal{D}\gamma_\alpha\int_{\Sigma} \extd^D x\; \Phi[\gamma_\alpha;x)\mathcal{C}\left[\gamma_\alpha,\frac{\delta}{\delta \gamma_\alpha};x\right)\Phi[\gamma_\alpha;x).
\een
Formally defining a ``functional functional" differential calculus through
\ben
\frac{\frak{d}\Phi[\gamma_\alpha;x)}{\frak{d}\Phi[\delta_\beta;y)}=\delta[\gamma_\alpha-\delta_\beta]\delta(x-y),
\een
the variation $\frak{d}S/\frak{d}\Phi$ would indeed reproduce (\ref{constreq}). However, such an explicit dependence on points seems very unnatural from the point of view of canonical quantum gravity, since points in $\Sigma$ are not variables on the phase space. An alternative and more attractive possibility arises once one realizes that in the canonical analysis of General Relativity one really has an integrated Hamiltonian involving non-dynamical quantities such as lapse and shift, which are originally part of the phase space, but can be removed from it after one has made sure that they decouple from the rest of the equations (see \eg \cite[sect. 4.4]{perttheory} on this). It seems therefore natural to integrate the constraint equations (\ref{constreq}) using Lagrange multipliers, and to extend $\Phi$ to be a functional of all original phase space variables, \ie both the dynamical variables $\gamma_\alpha$ and the Lagrange multipliers $\Lambda^i$. An action can then be defined as 
\ben
S=\int \mathcal{D}\gamma_\alpha \;\mathcal{D}\Lambda^i\; \Phi[\gamma_\alpha,\Lambda^i]\left(\int_\Sigma \extd^D x\;\Lambda^i(x)\mathcal{C}_i\left[\gamma_\alpha,\frac{\delta}{\delta \gamma_\alpha};x\right)\right)\Phi[\gamma_\alpha,\Lambda^i].
\label{genact}
\een
Variation with respect to the scalar field $\Phi$ then yields the equations of motion
\ben
0 = \mathcal{C}\left[\gamma_\alpha,\frac{\delta}{\delta \gamma_\alpha},\Lambda^i\right]\Phi[\gamma_\alpha,\Lambda^i]\equiv\left(\int_\Sigma \extd^D x\;\Lambda^i(x)\mathcal{C}_i\left[\gamma_\alpha,\frac{\delta}{\delta \gamma_\alpha};x\right)\right)\Phi[\gamma_\alpha,\Lambda^i].
\een
The integrated Hamiltonian constraint $\mathcal{C}$ is now just the Hamiltonian of canonical gravity; since this equation has to hold for arbitrary values of the Lagrange multipliers, it is equivalent to the form (\ref{constreq}) of the constraints. The fact that $\Lambda^i(x)$ were originally Lagrange multipliers is apparent from the action which does not contain functional derivatives with respect to these functions. So far the discussion is completely general and does not require a particular form of the constraints. Introducing smeared constraints is also what one does in loop quantum gravity since equations of the form (\ref{constreq}) are highly singular operator equations, as we already mentioned when discussing the Wheeler-DeWitt equation (\ref{wdw}). Notice that in this formulation we are treating {\it all} the constraints of canonical gravity on equal footing, and expect all of them to result from the equations of motion of the third quantized theory; this also means that, if we are working in metric variables, our field is defined on the space of metrics on $\Sigma$ before any constraint is imposed. 

An important point to note is that in the very idea of a continuum third quantization of gravity is implicit the use of a non-trivial kinetic term, corresponding to the {\it quantum} Hamiltonian (and possibly other) constraint(s). On the one hand this has to be immediately contrasted with the standard GFT action, an example of which we have given in the previous section, which has instead a trivial kinetic term (containing no derivatives in the connection or group representation). We will discuss more extensively this point in the concluding section. On the other hand, the fact that the {\it classical} action of the field theory on superspace encodes and makes use of the {\it quantum} constraint operators implies that choices of operator ordering and related issues enter prominently also at this level; this should clarify why the {\it classical} GFT action we have presented makes use, for example, of non-trivial $\star$-products in its very definition (which also end up encoding the canonical operator ordering chosen in the covariant path integral formulation of the discrete gravity dynamics at the level of the GFT Feynman amplitudes).
 
\

One more important issue is that of symmetries of the action (\ref{genact}). One should expect the symmetries of classical (and presumably quantum) General Relativity -- spatial diffeomorphisms, time reparametrizations etc. -- to be manifest as symmetries of (\ref{genact}) in some way. 

We would like to mention here that the kind of symmetries one normally considers, transformations acting on spacetime and fields defined on spacetime that do not explicitly depend on the dynamical fields, will from the viewpoint of the third quantized framework defined on some form of superspace appear as {\em global} rather than local symmetries; the parameters defining them do not depend on the point in superspace\footnote{In extensions of standard GR where one allows for topology change at the classical level by allowing degenerate frame fields, the situation may be different \cite{topchange}.}. This basic observation in our heuristic framework for a continuum third quantization is in agreement with recent results in GFT \cite{diffeogft}, where one can identify a global symmetry of the GFT action that has the interpretation of diffeomorphisms in simplicial gravity language.
 
\

If we now focus on formulations of General Relativity in connection variables, thus setting $\gamma_\alpha=\omega^{ab}_i$, the variable conjugate to the connection is generically some function of the frame field encoding the metric. In the representation of canonical quantum gravity wavefunctions as functionals of connections, it would be represented as a functional derivative. We would then follow the strategy outlined in the introduction for loop quantum gravity; we would redefine variables on the classical phase space to obtain a field which depends on holonomies only, and then rewrite the Hamiltonian in terms of holonomies and vector fields replacing the frame field variables. Issues of operator ordering and regularization will be dealt with in the way familiar from LQG. As a concrete example which will allow us to implement this procedure to some extent, as well as making connections to the GFT formalism we have presented earlier, we can specify to the case of 3d (Riemannian) general relativity in connection variables, which takes the form of a topological (BF) theory, heuristically defining the action
\ben
S=\int \mathcal{D} \omega^{ab}_{i}\;\mathcal{D} \chi^a\;\mathcal{D} \Omega^{ab}\; \Phi[\omega^{ab}_{i},\chi^a,\Omega^{ab}]H\Phi[\omega^{ab}_{i},\chi^a,\Omega^{ab}].
\label{superaction}
\een
In this case $H$ is taken to be the Hamiltonian of 3d general relativity (with vanishing cosmological constant),
\ben
H = -\frac{1}{2}\int_{\Sigma}\extd^2 x\left\{\epsilon_{abc}\chi^a\epsilon^{ij}R^{bc}_{ij}+\Omega^{ab}\nabla_j^{(\omega)}\pi^j_{ab}\right\},
\label{superham}
\een
involving $R^{bc}_{ij}$, the curvature of the $\frak{su}$(2) connection $\omega^{ab}_{i}$, the momentum $\pi^j_{bc}\sim\epsilon^{ij}\epsilon_{abc}e^a_i$ conjugate to $\omega^{ab}_{i}$, and the covariant derivative $\nabla^{(\omega)}$ associated to $\omega^{ab}_{i}$. We may view the Lagrange multiplier $\epsilon_{abc}\chi^a$ as an $\frak{su}$(2)-valued scalar. Instead of using the action (\ref{superaction}) one could also go to a reduced configuration space by dividing out $\SU(2)$ gauge transformations generated by the Gauss constraint $\nabla_j^{(\omega)}\pi^j_{ab}=0$, and define the action
\ben
S=\int \mathcal{D} \omega^{ab}_{i}\,\mathcal{D} \chi^a\; \Phi[\omega^{ab}_{i},\chi^a]\left(-\frac{1}{2}\int_{\Sigma}\extd^2 x\;\epsilon_{abc}\chi^a(x)\epsilon^{ij}R^{bc}_{ij}(x)\right)\Phi[\omega^{ab}_{i},\chi^a],
\label{superaction2}
\een
where $\Phi$ is now understood as a gauge-invariant functional of a connection and an $\frak{su}(2)$ scalar. At this formal level, an entirely analogous ``construction'' would be possible in metric variables, but, as we have mentioned, the rigorous construction of a Hilbert space for canonical GR in connection variables done in loop quantum gravity will allow us to make more progress towards a rigorous definition of the action if we work with an $\SU(2)$ connection as the basic variable of GR. Let us investigate this in more detail.

\sect{Towards a Rigorous Construction: Passing to Holonomies, Decomposing in Graphs}
\label{holonomy}

As we mentioned before, the use of connection variables allows us to use some of the technology developed for LQG (summarized in \cite{thomas}) to give a more rigorous meaning to the classical action (\ref{genact}). In the construction, one has to change phase space variables to holonomies, and decompose a generic functional on the space of (generalized) connections into functions only depending on a finite number of such holonomies. The definition of a measure on the space of (generalized) connections proceeds by defining a measure on a single copy of the gauge group (representing the space of holonomies), normalized to one, and then extending to arbitrary numbers of holonomies by so-called projective limits. This is essentially the same strategy also used to give a rigorous definition to the Wiener measure, see \cite{dewittmorette1}. A Hilbert space for canonical quantum gravity can then be constructed as the space of square integrable functionals on the space of (generalized) connections with respect to this measure. This strategy works very well for connections with compact gauge group. In our formulation, however, we require functionals that also depend on Lagrange multipliers, which are usually spatial scalars (or vectors); these are somewhat subtle to handle. One common way of dealing with scalars is by defining ``point-holonomies"
\ben
U_{\lambda, x}(\Lambda):=\exp(\lambda \Lambda(x))
\een
where we again view the argument of the exponential as an element of some Lie algebra $\frak{g}$ (for the action (\ref{superaction2}), $\frak{g}=\frak{su}$(2)). One can reconstruct $\Lambda(x)$ from its point holonomies, either by taking derivatives or taking the limit $\lambda\rightarrow 0$ (for a more extended discussion see \cite[sect. 12.2.2.2]{thomas}). Having introduced a parameter $\lambda$ into the construction, the Lagrange multiplier can now be treated on exactly the same footing as the connection, \ie defined in terms of its point holonomies, also $\SU(2)$ group elements for (\ref{superaction2}). We note that, although we will in the following focus on 3d Riemannian gravity, the same procedure would be applicable, with minor extensions, in any other setting where one has a gravity connection with compact gauge group, such as in 4d in time gauge.

Let us detail the construction step by step. Consider a graph $\Gamma$ embedded into the surface $\Sigma$, consisting of $E$ edges\footnote{We will not be too explicit about whether these are thought of as smooth, piecewise smooth, analytic etc. maps $[0,1]\mapsto \Sigma$, and refer to the literature for details.} and $V$ vertices. Take any square integrable function $f$ on $G^{E+V}$ and define a functional of a $\frak{g}$-connection $\omega$ and a $\frak{g}$-valued scalar $\chi$, called a {\em cylindrical function}, by\footnote{Note that the functions (\ref{cylindrical}) bear a close resemblance to the ``projected cylindrical functions" defined in \cite{projected} in an attempt to obtain a Lorentz covariant LQG formalism in 4d. Both are functions of both the gravitational connection and a Lagrange multiplier.}
\ben
\Phi_{\Gamma, f}[\omega^{ab}_{i},\chi^a] = f(h_{e_1}(\omega),\ldots,h_{e_E}(\omega),U_{\lambda,x_1}(\chi),\ldots,U_{\lambda,x_V}(\chi)).
\label{cylindrical}
\een
That is, the holonomies of the gravitational connection $\omega$ along the edges $e_i$ ($i=1,\ldots,E$) of the graph $\Gamma$ and the point holonomies of the Lagrange multiplier $\chi$ at the vertices $x_k$ ($k=1,\ldots,V$) are taken as arguments of $f$. Functional integration of such functions can be defined using the normalized measure on $G$,
\ben
\int \mathcal{D} \omega^{ab}_{i}\,\mathcal{D} \chi^a\; \Phi_{\Gamma, f}[\omega^{ab}_{i},\chi^a]:=\int \extd h_1\ldots \extd h_E\, \extd U_1\ldots \extd U_V\, f(h_1,\ldots,h_E,U_1,\ldots,U_V).
\label{integrate}
\een
The normalization of the measure is crucial for consistency, since one can think of $\Phi_{\Gamma, f}$ as associated to a larger graph $\Gamma'\supset\Gamma$ to which one extends $f$ trivially. Now the linear span of functionals of the form (\ref{cylindrical}) can be shown to be dense in the required space of functionals of generalized connections \cite{rovelli}. An orthonormal basis for the functions associated to each such graph, \ie an orthonormal basis for square integrable functions on $G^{E+V}$, is provided by the entries of the representation matrices for group elements in (irreducible) representations (due to the Peter-Weyl theorem \cite{peterweyl}),
\ben
T_{\Gamma,j_e,C_v,m_e,n_e,p_v,q_v}[\omega^{ab}_{i},\chi^a]\sim D^{j_1}_{m_1 n_1}(h_{e_1}(\omega))\ldots D^{j_E}_{m_E n_E}(h_{e_E}(\omega))D^{C_1}_{p_1 q_1}(U_{\lambda, x_1}(\chi))\ldots D^{C_V}_{p_V q_V}(U_{\lambda, x_V}(\chi)),
\label{inreps}
\een
where we omit normalization factors. Note that one associates one such representation $j_e$ to each edge and another one $C_v$ to each vertex in the graph.

If one is now interested in {\em gauge-invariant} functionals of the spin connection $\omega^{ab}_i$ and the field $\chi^a$, passing to a reduced phase space where the action of gauge transformations has been divided out, one discovers that a basis for these is provided by functions associated to generalized spin networks, graphs whose edges start and end in vertices, labeled by irreducible representations at the edges and vertices and intertwiners at the vertices projecting onto singlets of the gauge groups involved. In the case of 3d gravity considered in (\ref{superaction2}), there is only a single $\SU(2)$ under which connection and Lagrange multiplier transform, and schematically we would have\footnote{Since the sum is over an uncountably infinite index set, $\Phi(\Gamma,j_e,C_v,i_v)\neq 0$ for at most countably infinite $\{\Gamma,j_e,C_v,i_v\}$.} (an entirely analogous decomposition would have been possible at the level of cylindrical functions (\ref{cylindrical}), without going to spin networks)
\ben
\Phi[\omega^{ab}_{i},\chi^a] = \sum_{\{\Gamma,j_e,C_v,i_v\}}\Phi(\Gamma,j_e,C_v,i_v)\, T_{\Gamma,j_e,C_v,i_v}[\omega^{ab}_i,\chi^a]
\een
where $\Phi(\Gamma,j_e,C_v,i_v)\equiv \langle T_{\Gamma,j_e,C_v,i_v},\Phi\rangle$ in the kinematical LQG Hilbert space, and the spin-network functions $T_{\Gamma,j_e,C_v,i_v}$ are (again up to normalization factors)
\bea
T_{\Gamma,j_e,C_v,i_v}[\omega^{ab}_i,\chi^a] &\sim& D^{j_1}_{m_1 n_1}(h_{e_1}(\omega))\ldots D^{j_E}_{m_E n_E}(h_{e_E}(\omega))D^{C_1}_{p_1 q_1}(U_{\lambda, x_1}(\chi))\ldots D^{C_V}_{p_V q_V}(U_{\lambda, x_V}(\chi))\nn
\\&& (i_1)^{n_1 n_2\ldots q_1}_{m_1 m_2\ldots p_1}\ldots (i_V)^{q_V\ldots n_{E-1}n_E}_{p_V\ldots m_{E-1}m_E};
\eea
one takes the functions (\ref{inreps}) in the representations $j_e$ associated to the edges and the representations $C_v$ associated to the vertices, and contracts with intertwiners $i_v$ at the vertices. Pictorially we may represent a spin network as
\begin{figure}[htp]
\centering
  \begin{picture}(200,100)
  \put(20,50){\circle{10}}\put(23,47){$\bullet$}\put(2,47){$C_1$}\put(24,40){$i_1$}
  \bezier{125}(25,50)(50,85)(80,90)\put(80,95){\circle{10}}\put(78,87){$\bullet$}\put(63,93){$C_2$}\put(75,80){$i_2$}
  \bezier{125}(80,90)(120,90)(182,60)\put(186,60){\circle{10}}\put(179,57){$\bullet$}\put(193,55){$C_3$}\put(175,65){$i_3$}
  \bezier{241}(25,50)(100,65)(180,60)
  \bezier{125}(25,49)(70,35)(100,10)\put(100,5){\circle{10}}\put(98,7){$\bullet$}\put(83,3){$C_4$}\put(98,13){$i_4$}
  \bezier{241}(100,10)(140,10)(182,60)\bezier{467}(100,10)(175,0)(182,60)
  \bezier{315}(80,90)(95,80)(100,65)\bezier{315}(103,55)(120,15)(100,10)
  \put(30,70){$j_1$}\put(140,83){$j_2$}\put(110,53){$j_3$}\put(60,25){$j_4$}\put(135,25){$j_5$}\put(175,20){$j_6$}\put(100,35){$j_7$}
  \end{picture}
  \caption{A simple (generalized) spin network with vertices of valence 3 and 4.}
\end{figure}
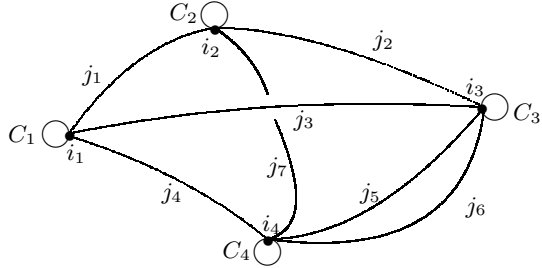

Using the expansion of the field $\Phi$ in spin-network functions, the action (\ref{superaction2}) can be formally written as
\ben
S=\sum_{s,s'\in S}\Phi(s)\Phi(s')\int \mathcal{D} \omega^{ab}_{i}\,\mathcal{D} \chi^a\; T_s[\omega^{ab}_{i},\chi^a]\left(-\frac{1}{2}\int_{\Sigma}\extd^2 x\;\epsilon_{abc}\chi^a(x)\epsilon^{ij}R^{bc}_{ij}(x)\right)T_{s'}[\omega^{ab}_{i},\chi^a],
\een
where we use the shorthand $s=\{\Gamma,j_e,C_v,i_v\}$ for the quantities determining a spin network, and $S$ is the (uncountably infinite) set of all such spin networks embedded into $\Sigma$. The task is then to determine the action of the Hamiltonian on a spin-network function and to use the orthogonality property $\langle T_s,T_{s'}\rangle=\delta_{s,s'}$ for appropriately normalized spin-network functions to collapse the sums to a single sum over spin networks. Let us again be more explicit on the details of the construction.

The Hamiltonian can be expressed in terms of holonomies by following the strategy used for 3d LQG in \cite{perez}: For given spin networks $s$ and $s'$, consider a cellular decomposition of the surface $\Sigma$, chosen so that $\Gamma$ and $\Gamma'$, the graphs used to define $s$ and $s'$, are contained in the union of 0-cells and 1-cells of the decomposition, into plaquettes $p$ with sides of coordinate length $\epsilon$ and approximate
\ben
\int_{\Sigma}\extd^2 x\;\epsilon_{abc}\chi^a(x)\epsilon^{ij}R^{bc}_{ij}(x)\approx\sum_p \epsilon^2 {\rm Tr}(\chi_p R_p[\omega])\approx 2\sum_p {\rm Tr}(\chi_p h_p[\omega])\approx \frac{2}{\lambda}\sum_p {\rm Tr}(U_{\lambda,p}(\chi) h_p[\omega])
\een
where $(\chi_p)_{bc}=\epsilon_{abc}\chi^a(x_p)$ for some point $x_p$ in $p$ which we approximate by the point holonomy, $(R_p)^{bc}=\epsilon^{ij}R^{bc}_{ij}$ which we approximate by the holonomy $h_p$ around the plaquette $p$:
\ben
U_{\lambda,p}=1+\lambda\chi_p+\ldots,\quad h_p=1+\half\epsilon^2 R_p+\ldots.
\een
Note that instead of treating $\lambda$ as an extra parameter taken to zero in a limit at the end, one could take $\lambda=\lambda(\epsilon)$ to be a suitable function of $\epsilon$ regularizing the operator in the then single limit $\epsilon\rightarrow 0$. 
The action becomes, formally,
\ben
S=\lim_{{\epsilon\rightarrow 0}\atop{\lambda\rightarrow 0}}\sum_{s,s'\in S}\Phi(s)\Phi(s')\int \mathcal{D} \omega^{ab}_{i}\,\mathcal{D} \chi^a\; T_s[\omega^{ab}_{i},\chi^a]\left(-\frac{1}{\lambda}\sum_p {\rm Tr}(U_{\lambda,p}(\chi) h_p[\omega])\right)T_{s'}[\omega^{ab}_{i},\chi^a].
\een
The next step is now to explicitly compute the action of the Hamiltonian, \ie the action of holonomies and point-holonomies, on spin-network functions. Since the latter form a basis of the space of gauge-invariant functionals of $\omega$ and $\chi$, one can expand
\ben
-\frac{1}{\lambda}\sum_p {\rm Tr}(U_{\lambda,p}(\chi) h_p[\omega]) T_{s'}[\omega^{ab}_{i},\chi^a] = \sum_{s''\in S} c^H_{s',s''} T_{s''}[\omega^{ab}_{i},\chi^a]
\een
Then one uses the orthogonality property for spin-network functions,
\bea
&&\int \mathcal{D} \omega^{ab}_{i}\,\mathcal{D} \chi^a \; T_s[\omega^{ab}_{i},\chi^a]T_{s''}[\omega^{ab}_{i},\chi^a]\nonumber
\\&\equiv &\int \extd h_1\ldots \extd h_E\, \extd U_1\ldots \extd U_V\, T_s[h_1,\ldots,h_E,U_1,\ldots,U_V]T_{s''}[h_1,\ldots,h_E,U_1,\ldots,U_V]\nonumber
\\& = &\delta_{s,s''},\eea
where functional integration over spin-network functions is defined as in (\ref{integrate}); one takes the union of the graphs $\Gamma$ and $\Gamma'$ used to define $s$ and $s'$ and extends the spin-network functions trivially to the union, integrating over $E+V$ copies of $\SU(2)$ (with appropriately normalized Haar measure), where $E$ is now the number of edges and $V$ the number of vertices in the union of $\Gamma$ and $\Gamma'$. The action now becomes
\ben
S=\lim_{{\epsilon\rightarrow 0}\atop{\lambda\rightarrow 0}}\sum_{s,s'\in S}\Phi(s)\Phi(s')c^H_{s',s},
\label{graphaction}
\een
with the action of the Hamiltonian now hidden in the matrix elements $c^H_{s',s}$, and one has to worry about the limits $\epsilon\rightarrow 0$ and $\lambda\rightarrow 0$. (In the usual formulation of LQG, where $\chi$ is not a phase space variable, the only limit is $\epsilon\rightarrow 0$, and one can prove \cite{perez} that the Hamiltonian is indeed a finite and well-defined operator in this limit.) 

In the traditional formulation of LQG \cite{rovelli, thomas}, the effect of the operator ${\rm Tr}(U_{\lambda,p}(\chi) h_p[\omega])$ on a spin network state is to add $p$ as a loop with $j=1/2$, a vertex at $x_p$ with $C_v=1/2$ and an intertwiner contracting the two $\SU(2)$ matrices. In this case we would have
\ben
-\frac{1}{\lambda}\sum_p {\rm Tr}(U_{\lambda,p}(\chi) h_p[\omega]) T_{s'}[\omega^{ab}_{i},\chi^a] = -\frac{1}{\lambda}\sum_p T_{s'\cup p_{j=C_{x_p}=1/2}}[\omega^{ab}_i,\chi^a] \equiv \sum_{s''\in S} c^H_{s',s''} T_{s''}[\omega^{ab}_{i},\chi^a]
\een
with $c^H_{s',s''}=-\frac{1}{\lambda}\sum_p \delta_{s'',s'\cup p_{j=C_{x_p}=1/2}}$, and hence
\ben
S=\lim_{{\epsilon\rightarrow 0}\atop{\lambda\rightarrow 0}}\sum_{s\in S}-\frac{1}{\lambda}\sum_p\Phi(s\cup p_{j=C_{x_p}=1/2})\Phi(s).
\label{superaction3}
\een
There have also been suggestions for non-graph-changing holonomy operators for loop quantum gravity which would mean that $c^H_{s',s''}=0$ unless $s'$ and $s''$ are defined on the same graph, so that one would have a (free) action ``local in graphs''. At this stage any attempt of a more concrete definition relies on input from a framework for canonical quantum gravity, in this case loop quantum gravity. 

Note that the expansion of the functional $\Phi$ into spin-network functions, imposed on us by the structure of the kinematical Hilbert space of LQG, suggests a slight change of perspective: While a supposed quantum field $\Phi[\omega^{ab}_{i},\chi^a]$ would be thought of as an operator creating a spatial hypersurface of topology $\Sigma$ with a continuous geometry described by $\omega$ and $\chi$, the elementary operators $\Phi(s)$ would create {\em graphs}, of arbitrary valence and complexity, with certain labels attached to them. One would have to contemplate a Fock space incorporating ``many-graph states", a massive extension of the ``one-graph" Hilbert space considered in LQG:
\ben
\frak{H}_{{\rm LQG}}\sim\bigoplus_{\Gamma}\frak{H}_{\Gamma}\quad\longrightarrow\quad\frak{H}_{{\rm 3rd\; quant.}}\sim \bigoplus_n \frak{H}_{{\rm LQG}}^n \sim \bigoplus_{\Gamma}\bigoplus_n \frak{H}_{\Gamma}^n,
\een
where each $\frak{H}_{{\rm LQG}}$ is already a space of functionals of connections.

This (we stress again, purely formal) picture in terms of graphs is of course closer to the discrete structures appearing in the interpretation of GFT Feynman graphs, and suggests to take the generalization of GFT to fields with arbitrary number of arguments seriously, as a vertex of arbitrary valence $n$ would be dual to an $n$-gon, and hence described by a GFT field with $n$ arguments. In the formulation (\ref{graphaction}), the graphs forming the spin networks in $S$ are still thought of as being embedded into a surface of topology $\Sigma$, but the next step towards a GFT-like interpretation in which one views them as abstract graphs is conceivably straightforward. A crucial difference between the continuum third quantization and GFT, however, is that between entire graphs represented by the operators $\Phi(s)$, and the building blocks of graphs represented by the GFT field, where entire graphs emerge as Feynman graphs in the quantum theory of the group field. We have stressed this point before, and will discuss its implications further in the next section.

If one wants to add an interaction term to the action, there is not much guidance from the canonical theory which can, from the interpretation of $\Phi$ as a quantum mechanical wavefunction, only ever supply a term quadratic in $\Phi$. Nevertheless, through the construction we have outlined the addition of such a term would in principle pose no additional difficulty. One would, in adding such a term, presumably impose conservation laws associated to the geometric interpretation of the quantities imposed, \eg impose conservation of the total area/volume represented by two graphs merged into one, which when expressed in terms of the canonically conjugate variable, the connection, would mean requiring ``locality" in the connection. This is indeed the strategy proposed for a minisuperspace model for third quantization in \cite{gftc}.

\section{Lessons for group field theories and Quantum Gravity, and some speculations}
Let us now conclude with a discussion of the relation between the two formulations of third quantized gravity we have presented. 

\

A first set of questions one may ask concerns the possibility of deriving a GFT model (more generally, the GFT setting) from the formal continuum third quantization framework we have presented. This means somehow discretizing the continuum third quantized action (\ref{graphaction}) to get a GFT action, or, better, identifying the restriction of the same which corresponds to a GFT dynamics.

The relevant questions are then: What is the limit in which a suitable GFT arises from (\ref{graphaction})? What is the expression for the same action if one adapts the Hamiltonian constraint to a generic triangulation, dual to the spin network graph? One possibility seems to be to make the assumption that $\Phi(s)=0$ except on a fixed graph $\gamma_0$, so that only certain modes are taken into account. One would then hope to be able to first identify the nontrivial contributions in the limit $\epsilon\rightarrow 0$, and then do a summation over the variables $C_v$ associated to Lagrange multipliers and get a GFT model. Following this line of thought, however, we immediately encounter a basic issue that sets the formal continuum third quantization and the GFT framework apart: whatever graph $\gamma_0$ is chosen, if it has to be one of those appearing in the decomposition of a gauge invariant functional of the connection coming from LQG, it will have to be a closed graph. A generic such graph, moreover, will be made of several vertices joined together by links, all without open ends. The fundamental GFT field, even in a combinatorial generalization of the GFT framework, will be typically associated to an open graph with a single vertex, and a certain number of outgoing links. Clearly, if a connection has to be sought along these lines a generalization of the GFT framework, or a re-writing of traditional LQG, is called for.
\

However, apart from such questions of the physical interpretation of spin networks in LQG, on general grounds not much faith can be placed on the third quantized formulation of gravity in the continuum, as a consistent, well-defined theory of quantum gravity. Thus the above question may have no more than an academic value, and simply be a heuristic goal. It is more sensible to view the problem in the opposite direction, and consider such a continuum formulation as an effective description of the dynamics of a more fundamental GFT in some approximation, one that captures in a convenient, if formal, way some relevant information about the GFT dynamics in a given physical regime.

\

It makes more sense, therefore, to ask a different set of questions, related to the big issue: what is the continuum approximation of the GFT dynamics? It may prove to be helpful to use the idea and general properties of the formal continuum third quantization for gravity to gain some insight on how the continuum limit of GFTs should be approached and on what it may look like.

\

Here we can immediately identify two main possibilities (not equally weighted):
 
\begin{itemize}
 
\item The first possibility is to try to define a continuum limit of GFTs already at level of classical GFT action, trying to obtain  from it the classical action of a third quantized field on continuum superspace.

This would mean defining some kind of limit of the GFT field, defined on a finite number of copies of the group manifold, producing a field over connection superspace, thus over an infinite dimensional manifold. More likely, one should expect to recover the generic wave function of continuum LQG, \ie a functional of a (generalized) connection that can be decomposed into an infinite combination of cylindrical functions, associated to all possible graphs embedded in a spatial manifold, each depending on a finite number of group elements. Clearly this will entail at the very least a generalization of the usual GFT framework to allow for all possible number of arguments of the GFT field(s), and combinatorial patters among them in the interaction term. We have already mentioned that this generalization is, in itself, not much of a problem within the GFT formalism. Still the step from usual GFT to this huge action involving all possible graphs will probably lead to the formulation of a GFT that will be very difficult to handle. Notice also that even this huge action would still codify the dynamics of (an infinite number of) {\it building blocks} of graphs, rather than graphs themselves, thus falling still a step shorter of the continuum third quantized action.

Next, along the same line of thought, let us discuss a few other more technical points. 

First, in introducing the formal third quantized action (\ref{superaction3}), we have assumed a graph-changing Hamiltonian; this type of dynamics, when reinterpreted from the point of view of a dynamics of graph {\it vertices}, implies a non-conservation of the same, and thus suggests naturally an underlying interacting field theory of such vertices, in which the canonical quantum dynamics is to be looked for in both the kinetic and interaction terms of an underlying (GFT) graph-vertex action, and the same interaction term that produces graph-changing also produces topology change. However natural, this way of encoding the graph dynamics is in sharp contrast  with the formal continuum one: The continuum third quantized action features a clear distinction between the topology-preserving contribution to the dynamics, corresponding to the canonical (quantum) gravitational dynamics, and encoded in a non-trivial kinetic term (Hamiltonian constraint), and the topology-changing contribution to it, encoded in the (generically non-local) interaction term. Notice that this distinction would remain true also in the case of a graph-preserving canonical Hamiltonian constraint. The only change, in the third quantized continuum action, would in fact be that the kinetic term would decompose (under the decomposition of the field on superspace into a sum of fields each associated to a (closed) graph)  into a sum of kinetic terms each associated to a different graph. Both the distinction  between topology-preserving and topology-changing contribution to the dynamics and a non-trivial kinetic term are not available in (standard) GFTs. Concerning the issue of a possible restriction of the GFT dynamics to a given (trivial) topology, there have been several encouraging results in the context of the GFT perturbation theory extending the large-$N$ limit of matrix models \cite{largen1,largen2}, but these clearly relate to the quantum dynamics of the GFT, and a similar approximation would not lead to any significant modification at the level of the GFT action, concerning the above issues. Still, the study of GFT perturbative renormalization \cite{GFTrenorm1,GFTrenorm2,GFTrenorm3,GFTrenorm4,GFTrenorm5,JosephValentin} is relevant because it may reveal (signs of this are already in \cite{JosephValentin}), that the very initial (microscopic) GFT action has to be extended to include non-trivial kinetic terms, in order to achieve renormalizability \footnote{Once more, the fact that in GFT one deals with an essentially standard QFT on a ``fixed background'', a group (or Lie algebra) manifold, indeed means that (almost) standard renormalization techniques from QFT are applicable.}. Another strategy that seems to lead to a solution to both the above-mentioned issues is the (mean field) expansion of the GFT action around a non-trivial background configuration. What one gets in doing so, quite generically, is in fact an effective dynamics for ``perturbations" characterized by a non-trivial kinetic term and thus a non-trivial topology- (and graph-)preserving dynamics, encoded in the linear part of the GFT equations of motion, and a topology- (and graph-)changing dynamics encoded in the non-linearities, \ie in the GFT interaction. This is the case before any continuum approximation is taken. This observation may suggest that the proper GFT dynamics which, in a continuum limit, would correspond to the (quantum) GR dynamics, is to be looked for in a different phase of the theory, and not around the trivial $\Phi=0$ (Fock, no space) vacuum. This point of view is supported by recent analysis of GFT models along these lines \cite{EteraWinston, noi, GFThydro, EffectiveHamiltGFT}.
\end{itemize}

Notwithstanding these considerations, it seems unlikely that a matching between discrete and continuum dynamics can be obtained working purely at the level of the action, simply because it is not obvious in what sense the GFT action should correspond simply to a discretization of the continuum third quantized action. As we have hinted at already, the discretization of the continuum third quantized framework represented by the GFT formalism seems to operate at a more radical, fundamental level, in terms of the very dynamical objects chosen to constitute a discrete quantum space: its microscopic building blocks themselves, rather than some discretization of it as a whole.
 
\begin{itemize} 
\item The second main possibility is that the continuum limit is to be defined at the level of the quantum theory, at the level of its transition amplitudes.

This seems much more likely, as a physical rather than purely formal approximation of the GFT dynamics. There are several reasons for this. One is simply the fact that it is the quantum GFT that defines the true dynamics of it, and that the classical action will be relevant only in some limited regime. A second one is the analogy with matrix models, for which we have seen in section \ref{matrmod} that the contact between discrete and continuum (third) quantization of gravity is made at the level of the transition amplitudes, or $n$-point functions of the theory. Moreover, we have seen that the continuum limit of the theory corresponds to a thermodynamic limit of the quantum/statistical theory and it is signaled by a phase transition. This is not something that can be seen at the level of the action alone.

A deeper implication of this perspective is that the emergence of a classical spacetime is the result of a purely quantum phenomenon, of the quantum properties of the underlying fundamental system. Once more, the continuum approximation is very different from the semi-classical approximation of the discrete quantum system.

When considering such a limit at the level of the quantum GFT theory, we must again bear several points in mind. Whatever path one decides to take, it should be clear, both from our presentation of the GFT formalism and of the third quantization idea in the continuum, that two main mathematical and conceptual issues have to be solved, in order to tackle successfully the issue of the continuum, both already mentioned above: the restriction to the dynamics of geometry for given (trivial) topology, and the step from a description of the dynamics of (gravitational) degrees of freedom associated to elementary building blocks of graphs, thus of a quantum space, to that of entire graphs, thus quantum states for the whole of space (as in LQG).
On the first issue, the recent results on the large-$N$ approximation of the GFT dynamics \cite{largen1,largen2} (for topological models) are certainly going to be crucial, and represent, even before any further understanding or development, a proof that the needed restriction can be achieved. The other example of a procedure giving, in passing, the same restriction is the mean field approximation of the GFT dynamics around a non-trivial background configuration. 

The second issue is, in a sense, more thorny, and does not seem to require only the ability to solve a mathematical problem, but rather some new conceptual ingredient, some new idea. So let us speculate on what these new ideas could be.
Looking back at the LQG set-up, one sees that the continuum nature of the theory, with its infinitely many degrees of freedom, is encoded in the fact that a generic wave function of the connection decomposes into a sum over cylindrical functions associated to arbitrarily complicated graphs. In terms of graph vertices, this means that the continuum nature of space is captured in the regime of the theory in which an arbitrary high number of such graph vertices, or GFT quanta, is interacting. The best way to interpret and use the GFT framework, in this respect, seems then to be that of many-particle and statistical physics. From this point of view, quantum space is a sort of weird condensed matter system, maybe a type of quantum fluid, where the continuum approximation would play the role of the hydrodynamic approximation in usual condensates or quantum fluids, and the GFT represents its microscopic quantum field theory, \ie the quantum theory of its ``atomic" building blocks. This perspective has been advocated already in \cite{GFTfluid} and it resonates with other ideas about spacetime as a condensate \cite{condensate1,condensate2} (see also \cite{GFThydro} for some recent results in this direction). In practice, the possibility that continuum spacetime physics is to be looked for in the regime in which (infinitely) many quantum building blocks, with their associated degrees of freedom, are interacting, leads immediately to the need for ideas and techniques from statistical field theory. In particular, it leads almost inevitably to the issue of possible phase transitions in GFTs, the idea being that the discrete-continuum transition is one such phase transition and that continuum spacetime physics (and General Relativity) are the resulting description of only one of the possible phases of the GFT system (it would be rather surprising, from experience in condensed matter theory, to find out that the GFT system, with its infinitely many interacting degrees of freedom, would possess only a single phase). 

It becomes even more apparent, then, that the study of GFT renormalization \cite{GFTrenorm1,GFTrenorm2,GFTrenorm3,GFTrenorm4,GFTrenorm5}, both perturbative and non-perturbative, will be crucial, in the longer run, for solving the problem of the continuum. The analogy with matrix models, where the above idea of discrete-continuum phase transition is indeed realized explicitly, and the generalization of ideas and tools developed for them to the GFT context, will be important as well.
 
Still, even in the case in which this turns out to be the correct way of approaching the problem, one mathematical/formal ingredient is missing to dwell in it more fully and fruitfully: a proper, complete particle representation of GFTs, \ie a rigorous Fock representation of GFT states (and thus of LQG states re-interpreted as many-particle states), clarifying the appropriate (generalized) statistics, and the translation for second quantized (LQG, simplicial gravity) operators to GFT ones, including those encoding the fundamental dynamics (LQG Hamiltonian constraint vs GFT action). This would be needed, in fact, for a more straightforward application of condensed matter ideas and results to GFTs. Work on this in in progress \cite{DanieleDerek}.
 
More generally, one has to be clear about what it is that one would hope to recover in the continuum limit -- would it be a continuum 3d (or higher-dimensional) gravity path integral? What exactly would one compare the result with, in order to check that one has correct limit? It seems very likely that one would have to go beyond a formal comparison and try to extract directly some physical prediction. Apart from the conceptual challenges of identifying what such a prediction could be, it seems clear for that one would need matter fields which also have to be incorporated into the GFT, making the task more difficult on a technical level too.

\end{itemize}

Probably a crucial role will be played by diffeomorphism invariance, as a guiding principle for recovering a good continuum limit and even for devising the appropriate procedure to do so. A first step towards elucidating the role of diffeomorphisms in GFT was taken in the recent work \cite{diffeogft} where a symmetry representing diffeomorphisms was identified in a ``colored'' version of the Boulatov model we have outlined in section \ref{gftsect}. This work as well as future work in this direction benefits from input from other approaches to quantum gravity which involve a discretized spacetime at some level, such as Regge calculus, where a lot of work has gone into identifying the role of diffeomorphisms in the discrete \cite{discdiffeo}. Similarly, insights about what sort of approximation of variables and which regime give the continuum can be looked for in classical and quantum Regge calculus. 

\

It is clear that in both cases above, the theory one would recover/define, starting from the GFT dynamics, would be a form of quantum gravitational dynamics, either encoded in a third quantized classical action, which would give equations of motion corresponding to canonical quantum gravity, or in a quantum gravity path integral. An extra step would then be needed to recover {\it classical} gravitational dynamics, presumably some modified form of General Relativity and of the Einstein equations, from it. This is where semi-classical approximations would be needed, in a GFT context. Their role will be that of ``de-quantizing'' the formalism, \ie to go from what would still be a third quantized formalism, albeit now in the continuum, to a second quantized one. More precisely, it would lead us from ``classical equations'' for the canonical wave function, obtained from the GFT classical dynamics, to the classical equations for the corresponding particle/universe, if applied directly at the level of the GFT action (in continuum approximation), or from the second quantized $n$-point functions (gravitational path integral) obtained as a restricted (``free particle'') sector of the GFT quantum dynamics (in the same continuum approximation) to a classical dynamics (Einstein's equations), if applied at the level of the GFT $n$-point functions.
This semi-classical approximation, which is, we stress once more, a very different approximation with respect to the continuum one, would be approached most naturally using WKB techniques or coherent states. On this last point it is worth distinguishing further between third quantized coherent states, coherent states for the quantized GFT field, within a Fock space construction for GFTs still to be defined properly, and second quantized ones, coherent states for the canonical wave function and defining points in the canonical classical phase space (\eg that of connectiondynamics), as defined and used in LQG. The last type of coherent states have been used to both extract an effective dynamics using mean field theory techniques and de-quantize the system, in one stroke, in \cite{GFThydro}. But even this recent work does not represent more than a first step in this direction.
This issue, more than those referring to the issue of the continuum limit, could also be fruitfully studied in a simplified mini-superspace third quantization \cite{gftc}; work on such mini-superspace third quantization, which can be seen both as a truncation of the GFT dynamics and as a generalization of the well-developed loop quantum cosmology \cite{lqcreview} to include topology change, is in progress. Besides possible phenomenological implications and the corresponding physical interest, this study could be useful to elucidate the relation between canonical (loop quantum gravity) and third quantized (GFT) dynamics, and the role of topology change. 

\section*{Acknowledgments}
This work is funded by the A. Von Humboldt Stiftung, through a Sofja Kovalevskaja Prize, which is gratefully acknowledged.

\end{document}